\documentclass[]{aastex}

\usepackage{emulateapj5,ifthen}
\shorttitle{Brightest Cluster Galaxies \& Intracluster Light}
\shortauthors{Lin \& Mohr}

\def\Om{$\Omega_M$\ }

\def\lCDM{$\Lambda$CDM\ }
\def\xray{\hbox{X--ray} }

\newcommand{\figtype}{EPS}
\def\myputfigure#1#2#3#4#5%
{\vskip#5pt\makebox[0pt]{\hskip#2in
\includegraphics[width=#3\textwidth]{#1}}\vskip#4pt\hfill}
\newenvironment{inlinefigure}{
\def\@captype{figure}
\noindent\begin{minipage}{0.999\linewidth}\begin{center}}
{\end{center}\end{minipage}\smallskip}

\begin{document}

\submitted{Submitted to ApJ May 30, 2004, Accepted August 30, 2004}

\title{$K$-band Properties of Galaxy Clusters and Groups: \\ 
    Brightest Cluster Galaxies and Intracluster Light}

\author{Yen-Ting Lin\altaffilmark{1} and Joseph J. Mohr\altaffilmark{1,2}}
\altaffiltext{1}{Department of Astronomy, University of Illinois,
Urbana, IL 61801; ylin2@astro.uiuc.edu}
\altaffiltext{2}{Department of Physics, University of Illinois,
Urbana, IL 61801; jmohr@uiuc.edu}

\begin{abstract}

We investigate the near-infrared $K$-band properties of the brightest cluster 
galaxies (BCGs) in a sample of 93 \xray galaxy clusters and groups, using data 
from the Two Micron All Sky Survey. Our cluster sample spans a factor of 70
in mass, making it sensitive to any cluster mass related trends. 
We derive the cumulative radial distribution for the BCGs in the ensemble, and
find that $70\%$ of the BCGs are centered in the cluster to within 5\% of the virial radius $r_{200}$; this
quantifies earlier findings that BCG position coincides with the
cluster center as defined by the \xray emission peak.
We study the correlations between the luminosity of the BCGs ($L_{b}$) and the mass and 
the luminosity of the host clusters, finding that BCGs in more massive clusters
are more luminous than their counterparts in less massive systems, and that
the BCGs become less important in the overall cluster light ($L_{200}$) as cluster mass increases.  
By examining a large sample of optically-selected groups we find that these 
correlations hold for galactic systems less massive than our clusters ($<3\times10^{13}M_\odot$).
From the differences between luminosity functions in high and low mass clusters,
we argue that BCGs grow in luminosity mainly by merging with other luminous 
galaxies as the host clusters grow hierarchically;  the decreasing BCG 
luminosity fraction ($L_{b}/L_{200}$) with cluster mass indicates that the rate of luminosity growth in BCGs is slow compared to the rate at which clusters acquire galaxy light from the field or other merging clusters.  

Utilizing the observed correlation between the cluster luminosity and mass and
a merger tree model for cluster formation, we estimate that the amount of intracluster light (ICL) increases with cluster mass; our calculations suggest that in $10^{15}M_\odot$ clusters more 
than 50\% of total stellar mass is in ICL, making the role of ICL very important
in the evolution and thermodynamic history of clusters.  
The cluster baryon fraction accounting for the ICL is in good agreement with the
value derived from Cosmic Microwave Background observations. 
The inclusion of ICL reduces the discrepancy between the observed cluster cold baryon 
fraction and that found in hydrodynamical 
simulations. Based on the observed iron abundance in the intracluster medium, we find that the 
ICL predicted by our model, together with the observed galaxy light, match
the iron mass-to-light ratio expected from simple stellar population models, 
provided the Salpeter initial mass function is adopted. The ICL also makes it easier to produce the ``iron excess'' found in the central regions of cool-core clusters.
\end{abstract}


\section{Introduction}
\label{sec:intro}

The most luminous cluster galaxies (hereafter referred to as brightest cluster
galaxies, BCGs) are a unique class of objects. They are
ultra-luminous ($\sim 10\,L_*$, where $L_*$ is the characteristic luminosity
in the galaxy luminosity function; e.g. \citealt{schombert86}) and huge in 
spatial 
extent \citep[effective radius $\sim$ 30 kpc; e.g.][]{schneider83b,schombert86,tonry87,
gonzalez00,gonzalez04}. They tend to lie very close to peaks of the cluster \xray emission \citep[][also \S\ref{sec:overview}]{jones84,beers86,rhee91},
and in velocity space they sit near the cluster rest frame 
\citep[e.g.][]{quintana82,zabludoff90,oegerle01};
the implication is that BCGs are located at the minimum in the cluster potential well. 
Some of the BCGs show multiple nuclei
\citep[e.g.][]{schneider83b,hoessel85,lauer88} and excess populations of globular clusters 
compared to normal early type galaxies.
\citep[e.g.][]{harris98}. Furthermore, correlations between the BCG luminosity
and various host cluster properties have been found \citep[e.g.][]{oemler76,
schombert87,edge91a,brough02}.  
All these properties indicate they may have a quite
unusual formation history compared to other elliptical galaxies.

There are generally three different models proposed for BCG formation: (1) Galactic
cannibalism-- massive galaxies gradually sink to the center of a cluster because
of dynamical friction, and the first galaxy arriving at the center grows in
luminosity and mass by merging with late-comers \citep[e.g.][]{ostriker75,hausman78}.
(2) Cooling flow-- stars formed out of the central cooling flow in clusters 
\citep[e.g.][]{cowie77}. 
(3) Rapid galactic merger during cluster collapse-- BCGs formed from
mergers between several galaxies that take place in groups or low mass clusters \citep{merritt85,dubinski98}. 
In this paper we use observed properties of a large sample of BCGs to 
examine their formation history.

Many of the first ranked galaxies have a large, diffuse envelope around them 
\citep[the ``cD'' galaxies; e.g.][]{tonry87,schombert88}. 
The facts that cD galaxies are only
found in clusters or local dense structures \citep{beers83} and that there 
exist weak correlations between envelope luminosity and some cluster global
properties \citep{schombert88} argue strongly that the formation of the 
envelopes must be closely connected to the cluster environment. 

Due to the dynamical processes such as the mean field tidal stripping 
\citep[e.g.][]{merritt84} and the cumulative
effects of impulsive encounters between galaxies or cluster substructures
(``galaxy harassment'', \citealt{moore96}), it is possible that stars may be
stripped from galaxies and orbit in intracluster space (which we refer to as the
intracluster stellar population or intracluster light, ICL). 
The investigation of diffuse light in intracluster space has a long history, 
first based on photographic plates \citep[][only to name a few]{zwicky51,welch71,oemler76,thuan77}.
Over the past 10 years, there have been many observational studies of the 
ICL made with CCDs; these include studies of tidal features
\citep[e.g.][]{gregg98,trentham98,calcaneo00}, planetary nebulae 
\citep[e.g.][]{arnaboldi02,feldmeier03}, globular clusters \citep[e.g.][]{jordan03}, 
red giant stars \citep{ferguson98,durrell02} and intracluster supernovae 
\citep{gal-yam03}.  Despite this wealth of data, the amount of ICL is very
hard to estimate; the fraction of total cluster light that is in ICL ranges
from less than $20\%$ to more than $50\%$ (this may partially reflect a
dependence of ICL fraction on cluster mass; see discussion in $\S$\ref{sec:icl}).
We note that a complicating factor in estimating the amount of the ICL is the 
possible
connection between the cD envelope and the ICL \citep{uson91}. Galactic haloes 
stripped from their parent galaxies may exist in the intracluster space in 
the form of a cD envelope \citep[e.g.][]{richstone76,malumuth84}.
In this paper we adopt the view that the cD envelope and the ICL are one and
the same, and thus do not attempt to make a distinction between the two 
(see \S\S \ref{sec:photo} \& \ref{sec:icllite} for further discussions).

This paper is the third in a series in which we systematically study the 
near-IR (NIR) properties of galaxy clusters. In \citet[][hereafter 
paper I]{lin03b} we develop basic tools for studying the correlation between 
total galaxy $K_s$-band luminosity and
cluster binding mass ($L$--$M$ relation). In \citet[][paper II hereafter]{lin04}
we show the tight $L$--$M$ relation in a
sample of 93 nearby clusters, using the data from the Two Micron All-Sky
Survey (2MASS, \citealt{jarrett00}). We find that the total number of
galaxies brighter than a luminosity cutoff is also well correlated with the 
cluster mass ($N$--$M$ relation), and the slope of the $N$--$M$ correlation
suggests that more massive clusters have smaller mean galaxy number per unit
mass, compared to the low mass clusters. The implication is that galaxies are
destroyed or stripped and lose light as the process of hierarchical structure formation proceeds toward higher mass scales.  In the present paper we aim to study the
correlation between the BCG luminosity and cluster mass, thus providing some
constraints on BCG formation scenarios, and investigate the amount of ICL that
may be present. This will allow us to probe the total cluster stellar
population.  Due to the depth of the 2MASS, it is not possible to directly measure
the amount of ICL; instead we estimate it from simple models that utilize 
the $L$--$M$ relation, the BCG light--cluster mass relation and models of hierarchical structure formation.

The plan of the paper is follows:
we describe in \S\ref{sec:overview} our cluster sample, the BCG selection 
criteria, and the photometry we use for BCG luminosity.
Our analysis of BCG properties is presented in \S\ref{sec:bcg}, where we
study the correlation between the cluster mass and the BCG luminosity, and the 
BCG-to-total galaxy luminosity fraction as a function of cluster mass 
(\S\ref{sec:bcglite}).  We discuss constraints on BCG formation scenarios 
(\S\ref{sec:bcgform}) and probe the properties of the 
second and third ranked galaxies (\S\ref{sec:m2m3}).  In \S\ref{sec:icl} we 
model the amount of ICL that may be present in clusters, and use it to examine 
the enrichment of the intracluster medium.  Finally we discuss and summarize 
our results in \S\ref{sec:disc} \& \S\ref{sec:summary}, respectively.

Throughout the paper we assume the density parameters for the matter and the
cosmological constant to be $\Omega_M = 0.3$, $\Omega_\Lambda = 0.7$,
respectively, and the Hubble parameter to be
$H_0=70\,h_{70}$~km~s$^{-1}$~Mpc$^{-1}$.

\begin{table*}[htb]
\begin{center}
\caption{BCG Data}
{\tiny
\begin{tabular}{lllcrrlllcrr}
\tableline \tableline
Cluster & $z$ & $T_X$ & BCG & $L_b^{iso}$\tablenotemark{\sharp} & $L_b^{ext}$\tablenotemark{\sharp} & Cluster & $z$ & $T_X$ &  BCG & $L_b^{iso}$\tablenotemark{\sharp} & $L_b^{ext}$\tablenotemark{\sharp} \\
Name     &     & (keV) & Name\tablenotemark{\dagger}  & $10^{11} L_\odot$ & $10^{11} L_\odot$ &   Name &     &   (keV) & Name\tablenotemark{\dagger}  & $10^{11} L_\odot$ & $10^{11} L_\odot$ \\
\tableline
a2319 & 0.0557 & 11.8 & 	 $19211004+4356443$ & 	$9.41\pm0.50$ &	$11.37\pm0.64$ &	a3562 & 0.0499 & 3.80 & 	 $13320334-3146430$ & 	$5.92\pm0.33$ &	$5.78\pm0.28$ \\
trian & 0.0510 & 9.50 & 	 $16381810-6421367$ & 	$17.50\pm0.89$ &	$18.87\pm1.25$ &	a2670 & 0.0762 & 3.73 & 	 $23541371-1025084$ & 	$9.78\pm0.72$ &	$12.31\pm1.08$ \\
a2029 & 0.0773 & 8.70 & 	 $15105610+0544416$ & 	$21.27\pm1.20$ &	$21.33\pm1.04$ &	a2657 & 0.0404 & 3.70 & 	 $23443040+0915499$ & 	$3.00\pm0.17$ &	$2.91\pm0.14$ \\
a2142 & 0.0899 & 8.68 & 	 $15582002+2714000$ & 	$6.06\pm0.57$ &	$6.67\pm0.63$ &	a1142 & 0.0349 & 3.7 & 	 $11005745+1030197$ & 	$5.09\pm0.22$ &	$4.70\pm0.19$ \\
a0754 & 0.0542 & 8.50 & 	 $09083238-0937470$ & 	$8.35\pm0.45$ &	$8.41\pm0.40$ &	a2634 & 0.0314 & 3.70 & 	 $23382938+2701526$ & 	$8.09\pm0.25$ &	$7.99\pm0.23$ \\
a1656 & 0.0232 & 8.21 & 	 $13000809+2758372$ & 	$10.10\pm0.19$ &	$9.56\pm0.17$ &	2a0335 & 0.0349 & 3.64 & 	 $03384056+0958119$ & 	$5.82\pm0.26$ &	$6.75\pm0.32$ \\
a2256 & 0.0601 & 7.51 & 	 $17042724+7838260$ & 	$7.59\pm0.51$ &	$7.77\pm0.46$ &	a3526 & 0.0114 & 3.54 & 	 $12484927-4118399$ & 	$8.05\pm0.13$ &	$7.75\pm0.17$ \\
a3667 & 0.0560 & 7.00 & 	 $20122726-5649363$ & 	$9.46\pm0.54$ &	$9.69\pm0.48$ &	a1367 & 0.0216 & 3.50 & 	 $11440217+1956593$ & 	$4.78\pm0.12$ &	$4.48\pm0.10$ \\
a2255 & 0.0806 & 6.87 & 	 $17122875+6403385$ & 	$6.43\pm0.47$ &	$7.03\pm0.54$ &	mkw03s & 0.0434 & 3.5 & 	 $15215187+0742319$ & 	$3.34\pm0.20$ &	$3.51\pm0.20$ \\
a0478 & 0.0881 & 6.84 & 	 $04132526+1027551$ & 	$8.61\pm0.44$ &	$10.81\pm0.65$ &	a1736 & 0.0458 & 3.50 & 	 $13272804-2719288$ & 	$9.84\pm0.40$ &	$9.63\pm0.35$ \\
a1650 & 0.0845 & 6.70 & 	 $12584149-0145410$ & 	$5.35\pm0.49$ &	$5.74\pm0.55$ &	hcg094 & 0.0417 & 3.45 & 	 $23171357+1842295$ & 	$10.63\pm0.41$ &	$10.31\pm0.41$ \\
a0644 & 0.0704 & 6.59 & 	 $08172559-0730455$ & 	$5.25\pm0.47$ &	$7.28\pm0.90$ &	a2589 & 0.0416 & 3.38 & 	 $23235741+1646379$ & 	$6.35\pm0.36$ &	$7.04\pm0.40$ \\
a0426 & 0.0183 & 6.33 & 	 $03194823+4130420$ & 	$7.54\pm0.14$ &	$8.17\pm0.29$ &	mkw08 & 0.0270 & 3.29 & 	 $14404287+0327555$ & 	$4.64\pm0.21$ &	$4.70\pm0.19$ \\
a1651 & 0.0860 & 6.30 & 	 $12592251-0411460$ & 	$8.90\pm0.70$ &	$10.82\pm0.96$ &	a4038 & 0.0283 & 3.15 & 	 $23474504-2808265$ & 	$4.07\pm0.20$ &	$4.82\pm0.27$ \\
a3266 & 0.0594 & 6.20 & 	 $04311330-6127114$ & 	$12.23\pm0.65$ &	$14.22\pm0.83$ &	a1060 & 0.0114 & 3.10 & 	 $10364282-2731420$ & 	$2.89\pm0.06$ &	$3.16\pm0.11$ \\
a0085 & 0.0551 & 6.10 & 	 $00415052-0918109$ & 	$11.53\pm0.53$ &	$11.24\pm0.46$ &	a2052 & 0.0348 & 3.10 & 	 $15164448+0701180$ & 	$6.59\pm0.31$ &	$7.52\pm0.38$ \\
a2420 & 0.0846 & 6.0 & 	 $22101878-1210141$ & 	$8.89\pm0.75$ &	$12.74\pm1.48$ &	a0548e & 0.0395 & 3.10 & 	 $05483835-2528404$ & 	$4.99\pm0.28$ &	$5.33\pm0.28$ \\
a0119 & 0.0440 & 5.80 & 	 $00561610-0115197$ & 	$8.45\pm0.39$ &	$8.55\pm0.38$ &	a2593 & 0.0433 & 3.1 & 	 $23242006+1438492$ & 	$6.58\pm0.40$ &	$7.48\pm0.45$ \\
a3391 & 0.0531 & 5.70 & 	 $06262045-5341358$ & 	$16.27\pm0.61$ &	$15.93\pm0.54$ &	a0539 & 0.0288 & 3.04 & 	 $05163713+0626526$ & 	$5.07\pm0.15$ &	$4.22\pm0.31$ \\
a3558 & 0.0480 & 5.70 & 	 $13275688-3129437$ & 	$13.50\pm0.63$ &	$14.95\pm0.72$ &	as1101 & 0.0580 & 3.00 & 	 $23135863-4243393$ & 	$5.42\pm0.38$ &	$7.80\pm0.70$ \\
a3158 & 0.0590 & 5.50 & 	 $03425295-5337526$ & 	$6.72\pm0.47$ &	$9.14\pm0.83$ &	a0779 & 0.0230 & 2.97 & 	 $09194687+3344594$ & 	$7.69\pm0.18$ &	$7.21\pm0.15$ \\
a1991 & 0.0590 & 5.4 & 	 $14543146+1838325$ & 	$5.73\pm0.40$ &	$6.85\pm0.48$ &	awm4 & 0.0326 & 2.92 & 	 $16045670+2355583$ & 	$7.11\pm0.24$ &	$7.05\pm0.22$ \\
a2065 & 0.0726 & 5.37 & 	 $15215562+2724530$ & 	$3.81\pm0.28$ &	$4.93\pm0.29$ &	exo0422 & 0.0390 & 2.90 & 	 $04255133-0833389$ & 	$5.14\pm0.25$ &	$5.13\pm0.21$ \\
a1795 & 0.0631 & 5.34 & 	 $13485251+2635348$ & 	$6.40\pm0.37$ &	$9.08\pm0.68$ &	a2626 & 0.0553 & 2.9 & 	 $23363057+2108498$ & 	$7.48\pm0.47$ &	$8.24\pm0.46$ \\
a3822 & 0.0760 & 5.12 & 	 $21550019-5739278$ & 	$6.62\pm0.45$ &	$6.86\pm0.47$ &	zw1615 & 0.0302 & 2.9 & 	 $16174059+3500154$ & 	$2.75\pm0.10$ &	$2.77\pm0.10$ \\
a2734 & 0.0620 & 5.07 & 	 $00112166-2851158$ & 	$6.23\pm0.42$ &	$7.70\pm0.55$ &	mkw09 & 0.0397 & 2.66 & 	 $15323201+0440516$ & 	$4.06\pm0.23$ &	$4.32\pm0.23$ \\
a3395sw & 0.0510 & 5.00 & 	 $06273625-5426577$ & 	$6.62\pm0.43$ &	$7.77\pm0.47$ &	a0194 & 0.0180 & 2.63 & 	 $01260057-0120424$ & 	$6.49\pm0.13$ &	$5.41\pm0.05$ \\
a0376 & 0.0484 & 5.0 & 	 $02460391+3654188$ & 	$5.62\pm0.33$ &	$6.60\pm0.41$ &	a0168 & 0.0450 & 2.6 & 	 $01145760+0025510$ & 	$5.26\pm0.26$ &	$5.54\pm0.26$ \\
a1314 & 0.0335 & 5.0 & 	 $11344932+4904388$ & 	$4.87\pm0.15$ &	$5.05\pm0.15$ &	a0400 & 0.0240 & 2.43 & 	 $02574155+0601371$ & 	$5.61\pm0.27$ &	$6.38\pm0.27$ \\
a2147 & 0.0351 & 4.91 & 	 $16021984+1620462$ & 	$4.70\pm0.17$ &	$4.71\pm0.16$ &	a0262 & 0.0161 & 2.41 & 	 $01524648+3609065$ & 	$4.25\pm0.11$ &	$4.08\pm0.10$ \\
a3112 & 0.0750 & 4.70 & 	 $03175766-4414175$ & 	$11.15\pm0.68$ &	$14.40\pm0.93$ &	a2151 & 0.0369 & 2.40 & 	 $16043575+1743172$ & 	$5.62\pm0.20$ &	$5.74\pm0.19$ \\
a1644 & 0.0474 & 4.70 & 	 $12571157-1724344$ & 	$11.35\pm0.61$ &	$12.67\pm0.69$ &	mkw04s & 0.0283 & 2.13 & 	 $12063891+2810272$ & 	$7.53\pm0.17$ &	$7.23\pm0.15$ \\
a2199 & 0.0303 & 4.50 & 	 $16283827+3933049$ & 	$8.45\pm0.28$ &	$8.05\pm0.24$ &	a3389 & 0.0265 & 2.1 & 	 $06222206-6456025$ & 	$5.64\pm0.20$ &	$5.32\pm0.17$ \\
a2107 & 0.0421 & 4.31 & 	 $15393904+2146579$ & 	$7.78\pm0.34$ &	$7.91\pm0.31$ &	ivzw038 & 0.0170 & 2.07 & 	 $01072493+3224452$ & 	$5.30\pm0.11$ &	$4.92\pm0.10$ \\
a0193 & 0.0486 & 4.2 & 	 $01250764+0841576$ & 	$7.77\pm0.31$ &	$8.21\pm0.34$ &	as0636 & 0.0116 & 2.06 & 	 $10302648-3521343$ & 	$3.49\pm0.07$ &	$3.15\pm0.06$ \\
a2063 & 0.0355 & 4.10 & 	 $15230530+0836330$ & 	$4.34\pm0.24$ &	$4.77\pm0.27$ &	a3581 & 0.0230 & 1.83 & 	 $14072978-2701043$ & 	$3.41\pm0.12$ &	$3.29\pm0.10$ \\
a4059 & 0.0475 & 4.10 & 	 $23570068-3445331$ & 	$10.41\pm0.50$ &	$10.75\pm0.50$ &	mkw04 & 0.0200 & 1.71 & 	 $12042705+0153456$ & 	$7.13\pm0.21$ &	$6.65\pm0.18$ \\
a1767 & 0.0701 & 4.1 & 	 $13360827+5912229$ & 	$8.67\pm0.56$ &	$11.29\pm0.86$ &	ngc6338 & 0.0282 & 1.69 & 	 $17152291+5724404$ & 	$6.07\pm0.18$ &	$5.70\pm0.15$ \\
a0576 & 0.0389 & 4.02 & 	 $07213023+5545416$ & 	$6.53\pm0.26$ &	$9.45\pm0.70$ &	a0076 & 0.0405 & 1.5 & 	 $00392632+0644028$ & 	$6.87\pm0.24$ &	$7.17\pm0.26$ \\
a3376 & 0.0456 & 4.00 & 	 $06004111-4002398$ & 	$5.81\pm0.34$ &	$5.96\pm0.31$ &	ngc6329 & 0.0276 & 1.45 & 	 $17141500+4341050$ & 	$4.00\pm0.13$ &	$3.61\pm0.10$ \\
a0133 & 0.0569 & 3.97 & 	 $01024177-2152557$ & 	$7.76\pm0.47$ &	$8.08\pm0.42$ &	as0805 & 0.0140 & 1.4 & 	 $18471814-6319521$ & 	$4.92\pm0.11$ &	$4.52\pm0.10$ \\
a0496 & 0.0328 & 3.91 & 	 $04333784-1315430$ & 	$7.00\pm0.28$ &	$7.35\pm0.28$ &	ngc0507 & 0.0165 & 1.26 & 	 $01233995+3315222$ & 	$5.62\pm0.12$ &	$5.45\pm0.12$ \\
a1185 & 0.0325 & 3.9 & 	 $11103843+2846038$ & 	$4.50\pm0.14$ &	$4.49\pm0.13$ &	ngc2563 & 0.0163 & 1.06 & 	 $08203567+2104042$ & 	$3.04\pm0.06$ &	$2.76\pm0.05$ \\
awm7 & 0.0172 & 3.90 & 	 $02542739+4134467$ & 	$6.79\pm0.14$ &	$6.39\pm0.15$ &	wp23 & 0.0087 & 1.0 & 	 $13165848-1638054$ & 	$2.87\pm0.05$ &	$2.95\pm0.10$ \\
a2440 & 0.0904 & 3.88 & 	 $22235694-0134593$ & 	$6.24\pm0.41$ &	$6.21\pm0.42$ &	ic4296 & 0.0133 & 0.95 & 	 $13363905-3357572$ & 	$8.13\pm0.16$ &	$7.42\pm0.14$ \\
a3560 & 0.0489 & 3.87 & 	 $13314673-3253401$ & 	$4.88\pm0.28$ &	$5.44\pm0.34$ &	hcg062 & 0.0137 & 0.87 & 	 $12530567-0912141$ & 	$2.99\pm0.09$ &	$2.77\pm0.08$ \\
a0780 & 0.0538 & 3.80 & 	 $09180565-1205439$ & 	$5.22\pm0.35$ &	$5.11\pm0.30$ &  & & & & \\
\tableline
\end{tabular}
}
\tablenotetext{\dagger}{The 2MASS designation, where the prefix 2MASXJ is omitted.}
\tablenotetext{\sharp}{$L_b^{iso}$ is obtained by subtracting 0.2 mag from the
  isophotal magnitudes; $L_b^{ext}$ is given by the extrapolated total 
  magnitudes. These are extinction and $k$-corrected luminosities.}
\end{center}
\vskip-35pt
\end{table*}

\section{Analysis Overview}
\label{sec:overview}

In this section we discuss our cluster sample, the BCG selection, the method we 
use to estimate total cluster luminosity, and the photometry in
the 2MASS catalog that we adopt to obtain BCG luminosity. We further test how
reliable the chosen photometry is in representing the total galaxy light.

\subsection{Cluster Sample and Total $K$-band Luminosity}
\label{sec:sample}

\begin{inlinefigure}
   \ifthenelse{\equal{\figtype}{EPS}}{
   \begin{center}
   \epsfxsize=8.cm
   \begin{minipage}{\epsfxsize}\epsffile{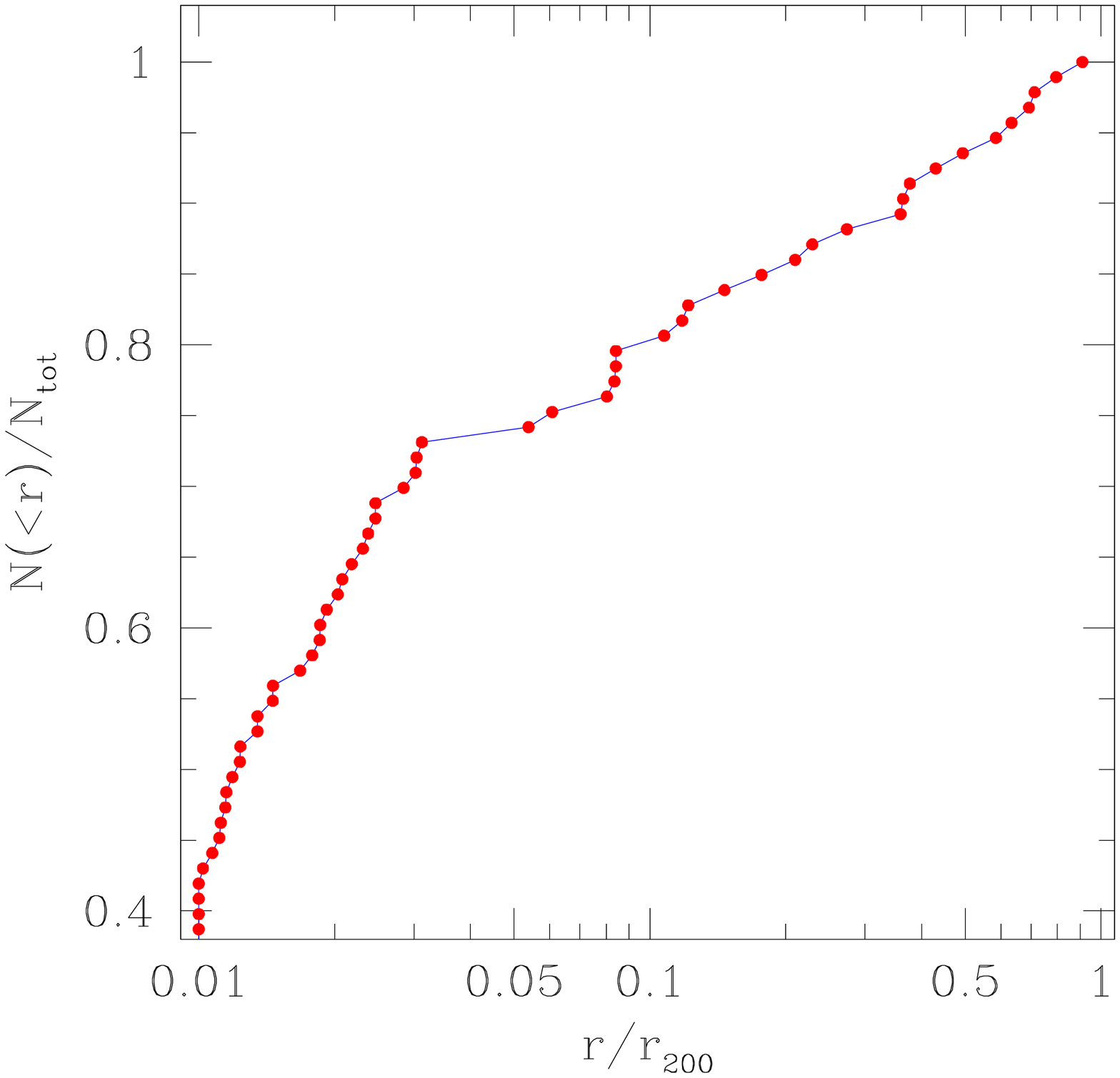}\end{minipage}
   \end{center}}
   {\myputfigure{f1.pdf}{0.0}{1.1}{-90}{-30}}
   \figcaption{\label{fig:dist}
	Cumulative projected radial distribution of the BCGs. The distance is 
	normalized to the virial radius ($r_{200}$) of each cluster. The number
	of galaxies located within a given distance is 
	normalized to the total BCG number $N_{tot} = 93$. Notice that the BCGs 
	located within 1\% of the viral radius are shown to be at
	$r/r_{200}=0.01$. Note that 75\% of the BCGs are located within 
	$0.06r_{200}$.
     }
\end{inlinefigure}

Our cluster sample is built from several \xray cluster catalogs; please refer to
\S 2 of paper II for the catalog references and a more detailed explanation of 
our analysis.  Here we give a brief account of our procedures.
Due to the depth of the 2MASS, we study clusters only out to $z \le 0.09$; we 
restrict the clusters to lie above the Galactic plane ($|b|>10^\circ$).
All our clusters have \xray emission-weighted mean temperature $T_X$ 
measurements, from which we estimate the virial radius $r_{200}$, within which
the enclosed mean overdensity is 200 times the critical density of the universe,
$\rho_c$.
The cluster masses range from $3\times 10^{13} h_{70}^{-1} M_\odot$ to
$2\times10^{15} h_{70}^{-1} M_\odot$, spanning approximately a factor of 70 in virial mass.
We define the cluster center as the peak of the \xray emission, either from the
cluster catalogs or from archival {\it ROSAT} images. The cluster redshift is
from the catalogs or NED/SIMBAD. For each cluster, we build a galaxy catalog 
from the 2MASS extended source catalog for galaxies brighter than the 2MASS
completeness limit $K_s = 13.5$ mag (hereafter we denote $K_s$ by $K$ for 
simplicity) that lie within the virial radius.

Within each galaxy catalog, we sort the galaxies according to their apparent
magnitudes, and search for redshift information from the NED for the bright 
galaxies. Redshift measurements are available for
most of the galaxies.  We consider galaxies with recession velocity within 
$v_c \pm 3\sigma_v$ as cluster members, where
$v_c$ and $\sigma_v$ are cluster recession velocity and velocity dispersion, 
respectively, which are collected from the literature 
\citep{struble99,wu99}. We do not choose BCGs based on their morphology.

We list in Table 1 the BCGs in our sample. The table contains the cluster name,
cluster redshift and \xray temperature, BCG name (under the 2MASS designation),
BCG luminosities derived from 2MASS isophotal and extrapolated total 
photometries, respectively (see \S\ref{sec:photo}).

There is one cluster whose BCG could not be unambiguously determined, because 
no redshift measurement is available (Exo 0422-086);
we assign the brightest galaxy as the BCG, as it lies very
close to the cluster center (the projected distance is $0.007\,r_{200}$). 
Removing this system from the sample does not change our results.

In Fig~\ref{fig:dist} we show the cumulative radial distribution of the BCGs,
with radial distance normalized to the virial radius of the host clusters.  The
majority of the galaxies that we assign as BCGs is located at the center of the
cluster potential, as traced by the \xray emission. For example, about 45\% of
the BCGs are located within 1\% of the virial radius, and 90\% within $0.38\,
r_{200}$.  It has long been recognized that BCGs lie very close to the cluster center  \citep[e.g.][]{jones84,beers86,rhee91,postman95,lazzati98}; what is
new from our analysis are the large cluster sample size and the adoption of the cluster
virial radius as a fundamental scale. An important implication
is that, in the absence of \xray images or detailed kinematic information about
the galaxies, the BCG position can serve as a good proxy for the cluster
center.

Except for those clusters undergoing major mergers, in which case BCGs may be
far from the minimum of the cluster potential, we do not expect BCGs to lie far
from the \xray peaks.  Within this context, the figure also allows us to 
estimate the accuracy of our BCG selection.
 In cases of merging, the galaxy we identify as the BCG may well be
a chance interloper. From the figure we see that $\sim 8\%$ of our BCGs lie at
$r>0.5 r_{200}$; indeed, most of the second-ranked galaxies of these clusters
lie at $r<0.1 r_{200}$.  This implies a $\lesssim 10\%$ contamination rate from
non-BCGs in the BCG luminosity--cluster mass correlations (\S\ref{sec:bcg}),
and this may be a source of the scatter. We note, however, restricting the BCGs
to lie well within the cluster virial radius (e.g. $r_{500} \approx 0.6 
r_{200}$, within which the mean density is $500\rho_c$) has very little effect
on the cluster $L$--$M$ relation, or the BCG light fraction (see 
\S\ref{sec:bcglite} below). 

Given the high frequency of clusters exhibiting X--ray evidence of mergers ($50-70\%$, as inferred
from the centroid variations, e.g. \citealt{mohr95}), the preponderance of centrally located
BCGs in this sample (Fig~\ref{fig:dist}) seems rather surprising.  This may
suggest that the timescale for the BCG to sink to the cluster potential minimum is short compared to  the relaxation timescale of the intracluster gas.  Alternatively, the scale of merger required to perturb the X-ray properties of the cluster could be smaller than the scale of the merger required to perturb the BCG from the cluster center.

To estimate the total cluster light for each cluster, we calculate the total 
observed number and luminosity from member galaxies (excluding the BCG);
corrections for background/foreground 
galaxies are estimated statistically. From the total number and 
luminosity we solve for the cluster luminosity function (LF) $\phi(L)$. 
Specifically, with a fixed faint-end slope $\alpha=-1.1$, we solve 
for the characteristic luminosity  $L_*$ and density $\phi_*$ in the functional form proposed by 
\citet{schechter76}.  We note the choice of $\alpha$ is consistent with the 
``stacked'' luminosity function formed from all our 93 clusters (see \S 3.2 in
paper II).  The total luminosity from galaxies, which are within the 
virial region and are brighter than a minimum luminosity
$L_{min}$, is then $L_{200} = \int_{L_{min}}^\infty \phi(L) L dL + L_{b}$, 
where $L_{b}$ is the BCG luminosity, and we choose $L_{min}$ that corresponds
to $M_{min} = -21$ mag. Our approach is described in more detail in \S 2 of 
paper I.

\subsection{2MASS Photometry of BCGs}
\label{sec:photo}

We use the $\mu_K = 20$ mag/arcsec$^2$ isophotal elliptical aperture magnitudes 
provided by the
2MASS extended source catalog for both the BCGs and other galaxies. 
The effects of Galactic extinction and $k$-correction are included.
We use the value of the extinction coefficient provided by the NED at the 
cluster center, and adopt a $k$-correction of the form $k(z) = -6\log (1+z)$ 
\citep{kochanek01}.  More importantly, we 
make a correction of 0.2 mag to convert to the ``total'' magnitudes, following
\citet{kochanek01}. This value is inferred by comparing 2MASS (second
incremental release) isophotal magnitudes with deeper photometry 
\citep{kochanek01}, and is consistent with the estimate that the isophotal 
magnitudes capture $80-90\%$ of the integrated flux for normal galaxies
\footnote{http://www.ipac.caltech.edu/2mass/releases/allsky/doc/explsup.html}.
However, given the peculiar properties of the BCGs (especially those that belong
to the cD class), we do not expect the simple correction for normal galaxies to
apply to BCGs. Below we use published BCG photometry to investigate how well this naive correction works in recovering the BCG ``total'' light. We emphasize here that we do {\it not} 
attempt to {\it measure} the total BCG light, but are mostly concerned with the correlations
between an objective measure of BCG light and cluster mass (\S\ref{sec:bcg}).

Bearing our objective in mind, it is crucial that the isophotal photometry we 
choose does not introduce any cluster mass related systematics. To be specific, we have to understand if
there is any (cluster) mass-dependent structural variation in the BCGs that will
affect the isophotal photometry. To this end, we gather from the literature
observed surface brightness profiles for 49 BCGs whose host clusters have
measured \xray temperatures (primarily from 
\citealt{graham96}, with additional measurements from 
\citealt{gonzalez00,fasano02,graham02}; below they are referred to as the known
profile BCG sample). With appropriate colors $R-K=2.6$,
$r-K=2.5$, $I-K=2.0$ \& $B-K=4.2$ for early type galaxies, we transform the 
measured profiles into the $K$-band, and calculate the differences between 
the $\mu_K = 20$ mag/arcsec$^2$ magnitudes and those corresponding to $\mu_K = 
20.9$ \& 21.5 mag/arcsec$^2$. We caution that because the completeness
limit of \citet{graham96} sample is at $\mu_R = 23.5$ mag/arcsec$^2$, some of 
the magnitude differences shown in Fig~\ref{fig:mdiff} are {\it extrapolations}.
The resulting 
differences depend on the surface brightness profiles of the galaxies, as well 
as the surface brightness at which the magnitudes are obtained. From the figure
it is seen that on average the magnitude differences are $\overline{\Delta m} (\mu_{20}-
\mu_{20.9}) = 0.26$, $\overline{\Delta m} (\mu_{20}-\mu_{21.5}) = 0.43$.
Furthermore, we notice that the BCG
surface brightness profile are typically described by either a Sersic profile or
a power-law \citep{graham96}. The average magnitude differences for those 
galaxies best-fit by
Sersic profiles (36 galaxies) are 0.21 \& 0.36 mag at different surface 
brightness, while for the 13 galaxies best described by
power-laws they are 0.38 \& 0.63. The larger magnitude difference for the
latter is expected, because of the flatness of the profiles. This exercise also
indicates that the surface brightness profiles must steepen at larger radii
(or fainter surface brightness) or the ``total'' light of the BCGs would be ill-defined.

Interestingly, we note
the differences do not show any obvious trend with either cluster redshift or
temperature (i.e. mass). As the 49 BCGs span wide ranges in the best-fit 
profile parameter space (c.f. \citealt{graham96}) and their host clusters in the
$z-T_X$ space, the results of this test should be representative of
nearby BCGs and, therefore, our galaxy sample.

The above test shows that our simple scheme of ``correcting'' the galaxy 
magnitudes by subtracting 0.2 mag from the isophotal magnitudes should not 
introduce any systematic correlation between the BCG light and cluster mass. 
However, it is also important to see how much light is recovered from this
0.2 mag correction. Apparently this depends on the BCG light profile and the
depth desired.
2MASS provides photometries derived from larger apertures, such as Kron
and ``total'' (extrapolated) magnitudes. These are obtained by fitting a
Sersic profile in a large area around each galaxy. We choose to use the 
isophotal magnitudes for the BCGs, as used for other galaxies, for the
homegeneity of the analysis, and also to prevent possible contamination from
nearby stars or surface brightness fluctuations on the sky.

We examine the differences between the isophotal and the total extrapolated
magnitudes for the BCGs, and find that for about 75\% of all 93 BCGs
the magnitude difference is 0.2 mag, with a scatter of $\sim0.1$ mag. This implies a 0.2 mag correction
to the isophotal magnitude gives an accuracy good to $\sim 10\%$. Most of the remaining
25\% of the BCGs (23 galaxies) belong to clusters at $0.02\le z\le 
0.09$ that have masses
$2\times 10^{14} M_\odot \le M_{200} \le 7.5\times 10^{14} M_\odot$.
By inspecting the postage stamp images of these 
galaxies\footnote{http://irsa.ipac.caltech.edu/applications/2MASS/PubGalPS/}, we
find that most of them have nearby bright sources at $0\arcmin.5-1\arcmin$ 
around them.

\begin{inlinefigure}
   \ifthenelse{\equal{\figtype}{EPS}}{
   \begin{center}
   \epsfxsize=8.cm
   \begin{minipage}{\epsfxsize}\epsffile{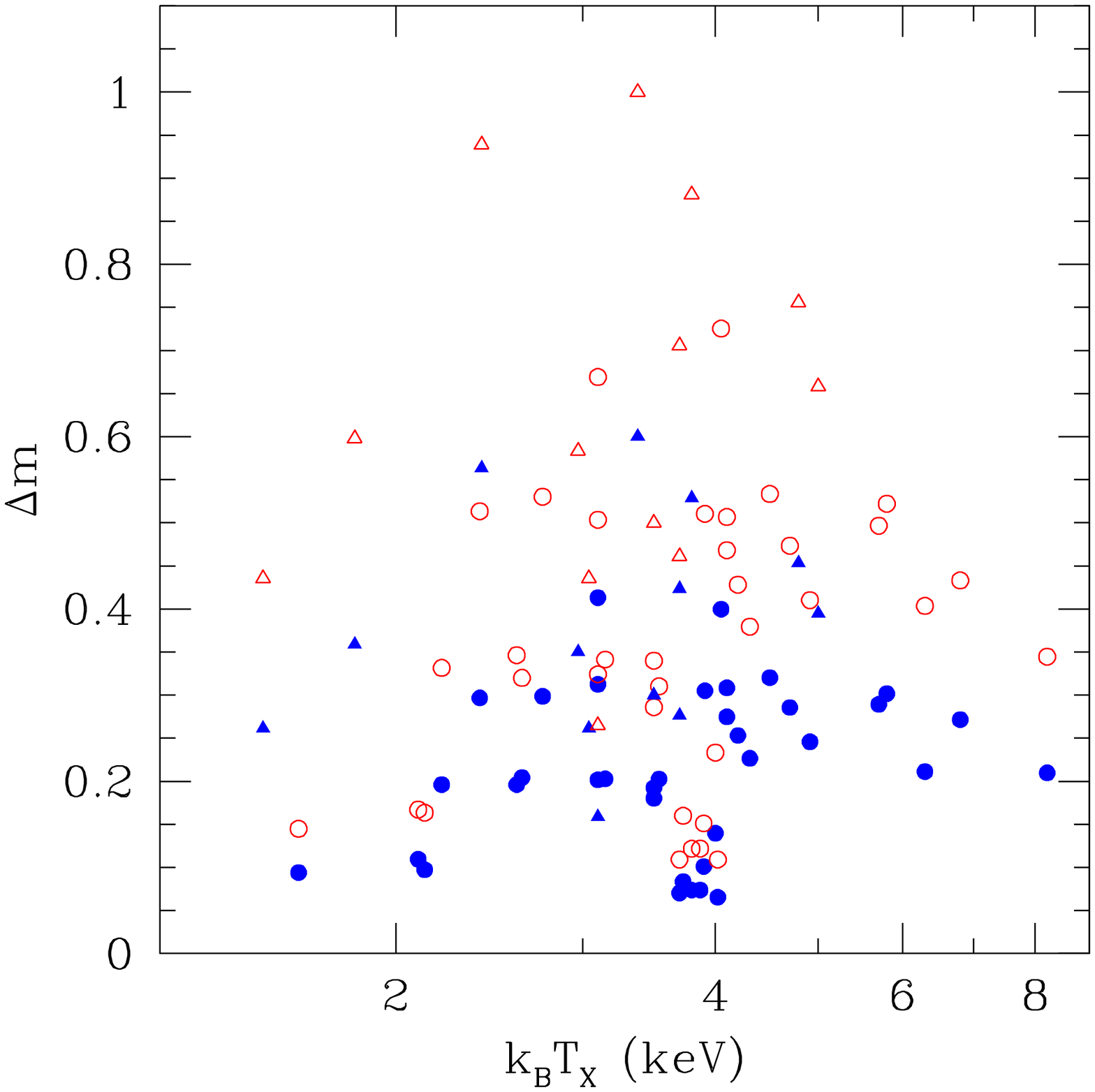}\end{minipage}
   \end{center}}
   {\myputfigure{f2.pdf}{0.0}{1.1}{-90}{-30}}
   \figcaption{\label{fig:mdiff}
	Differences in isophotal magnitudes at $\mu_K=20$ mag/arcsec$^2$ and at
	other surface brightness (solid points: 20.9 mag/arcsec$^2$, hollow 
	points: 21.5 mag/arcsec$^2$). The circular points indicate the
	galaxy is best-fit by a Sersic profile, while the triangles
	represent those best-fit by a power-law surface brightness profile.
	Notice that there is no apparent trend between the magnitude differences
	and cluster mass (\xray temperature).
     }
\end{inlinefigure}

In addition to possible contamination from nearby sources, 
cosmic dimming is also a probable cause of the larger difference between the 
photometries.  At greater 
redshifts the cosmic dimming may affect the isophotal photometry and in turn any
correlation between the BCG and cluster mass, as more massive clusters in our 
sample tend to lie at higher redshifts. We artificially place the galaxies
in the known profile BCG sample
at different redshifts and evaluate the magnitudes of cosmic dimming.
It is found that if the clusters were all at $z\sim 0.1$, dimming would cause 
a change of 0.05 to $\lesssim 0.2$ mag. 
Furthermore, we do not see any correlation between
the profile shape and the redshift for the BCGs of known profiles.
In particular, we note that 7 of the 23 known profile BCGs that lie in our sample of 93 
have no apparent systematic differences in their
profile parameters compared to the other BCGs of known profiles.

Thus, for the analysis below we estimate the BCG (as well as other cluster
galaxies) magnitude as 0.2~mag less than the 2MASS elliptical, isophotal 
magnitudes.

\section{Connection between BCGs and Host Clusters}
\label{sec:bcg}

In paper II we analyze the
$L$--$M$ relation, the correlation between the total light from all the cluster 
galaxies and the cluster mass. Let $L_g$ denote the luminosity from
all galaxies except for the BCG; we find that $L_{200} = L_g + L_{b} \propto
M_{200}^{0.72\pm 0.04}$, while $L_g \propto M_{200}^{0.82\pm 0.04}$. This
implies that the importance of the BCG contribution to cluster light depends on 
the cluster mass, and that BCG luminosity is very important in the cluster light
budget.  By examining any correlations between properties of the BCGs and host
clusters, one may be able to place constraints on the formation scenarios 
of the BCGs and more broadly, the evolution of the clusters.

Below we first study the $L_{b}$--$M_{200}$ correlation in detail, and then 
discuss
the BCG contributions to the total light across a factor of 70 in cluster mass
(\S\ref{sec:bcglite}).
We then use these observations to discuss the BCG formation history
(\S\ref{sec:bcgform}).
Finally we investigate the photometric and kinetic properties of the second-- 
and third--ranked galaxies (hereafter G2 and G3) in a subset of our cluster 
sample (\S\ref{sec:m2m3}).

A word of caution is in order here-- although in this section we often refer to 
$L_{200}$ as the total light in a system, it actually means the total light in
galactic objects. As we argue in \S\ref{sec:intro}, the evidence for a stellar
population in intracluster space is firm, and the ICL contribution to the total
cluster light may not be negligible. Furthermore, we regard stars in 
extended diffuse envelopes around some BCGs as part of the ICL.
We therefore use the notation $L_{tot}$ to denote light from all known stellar
populations in clusters (the BCG and the ICL, and the galaxies) brighter than
a luminosity cutoff.  

\subsection{BCG Luminosity and Cluster Mass}
\label{sec:bcglite}

\begin{inlinefigure}
   \ifthenelse{\equal{\figtype}{EPS}}{
   \begin{center}
   \epsfxsize=8.cm
   \begin{minipage}{\epsfxsize}\epsffile{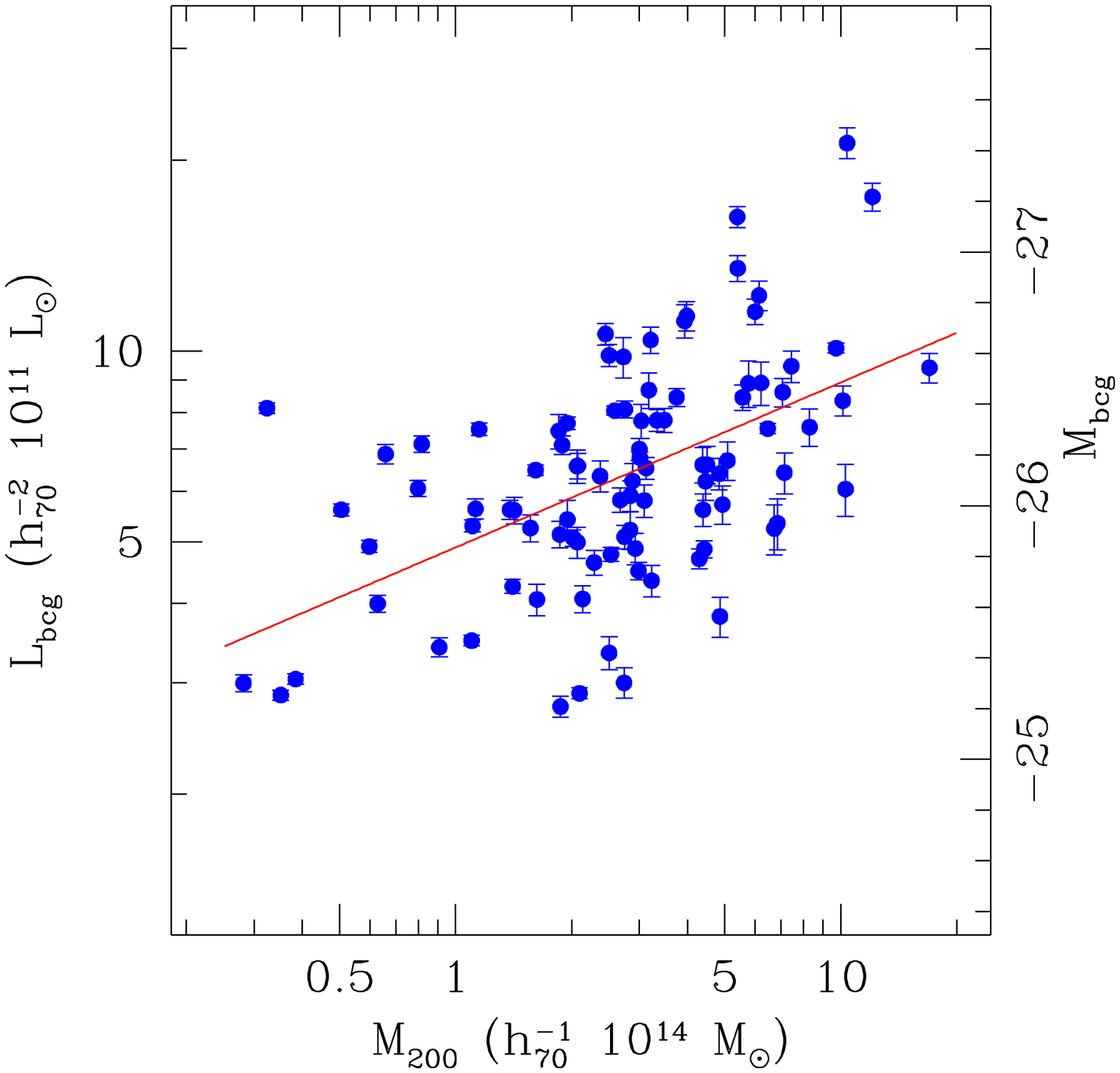}\end{minipage}
   \end{center}}
   {\myputfigure{f3.pdf}{0.0}{1.1}{-90}{-30}}
   \figcaption{\label{fig:bcglm}
	The correlation between the BCG luminosity (estimated from the isophotal
	magnitudes) and the cluster virial mass. On the right axis is shown the
	corresponding magnitudes.  Overall the BCG luminosity scales with 
	cluster mass as $L_b \propto M_{200}^{0.26\pm0.04}$ (the solid line).
     }
\end{inlinefigure}

BCGs are well-known for the similarity of their luminosity \citep[e.g.][]{
sandage72,sandage76,hoessel80,schneider83a,collins98}, and there have even been attempts to use them as standard candles in a cosmological context \citep[][]{postman95}. In 
Fig~\ref{fig:bcglm} we plot the BCG $K$-band isophotal magnitudes 
against the mass of their parent cluster. The BCGs in our sample have a mean 
magnitude of $M_K = -26.12 \pm 0.05$ mag, with a 0.44 mag scatter about the 
mean. For comparison, the characteristic magnitude of the composite cluster LF 
for our cluster sample is $M_{K*} =  -24.34$ (paper II).

Considering the vastly different environments in clusters in our sample,
it is not surprising that a spread in BCG luminosity is present; however, more 
interesting is the apparent
correlation between $L_{b}$ and cluster binding mass. Below the mass scale
$\sim 10^{14} M_\odot$, $L_{b}$ seems to be rather constant (although see 
\S\ref{sec:usgc}), however, 
for systems more massive, there is a trend; a least-square fit to the data for
11 systems with $M_{200}<10^{14} M_\odot$ gives
\[
\frac{L_b}{10^{11} h_{70}^{-2} L_\odot} = 5.7^{+3.4}_{-2.1} \left( \frac{M_{200}}{10^{14}  h_{70}^{-1}M_\odot} \right)^{0.32\pm0.30},
\]
consistent with a flat distribution;
for 82 clusters more massive than $10^{14} M_\odot$ we have
\[
\frac{L_b}{10^{11}  h_{70}^{-2}L_\odot} = 4.4\pm0.3 \left( \frac{M_{200}}{10^{14} h_{70}^{-1} M_\odot} \right)^{0.33\pm0.06}.
\]
When fitting the whole BCG sample, we find 
\begin{equation}
\label{eq:bcglm}
\frac{L_b}{10^{11}  h_{70}^{-2}L_\odot} = 4.9\pm0.2 \left( \frac{M_{200}}{10^{14} h_{70}^{-1} M_\odot} \right)^{0.26\pm0.04}.
\end{equation}
The Spearman correlation coefficient is 0.51, with a probability of $1.5\times 
10^{-7}$ that such correlation happens by chance. The fractional scatter about 
the fit is 34\%.

There have been many attempts to examine how well BCGs correlate with the
parent clusters, as this may reveal the connection between the two
\citep[e.g.][among
others]{sandage73,sandage76,hoessel80,schneider83a,postman95,graham96,collins98,
katayama03}.  Using photographic data, \citet{oemler76} shows both $L_{b}$ \&
$L_{b+env}$ correlate with the total cluster luminosity for a sample of cD galaxies, where
$L_{b+env}$ is the luminosity of the galaxy and the cD envelope associated with
it. This result is
confirmed later within a much larger sample \citep{schombert87,schombert88}.
The former study also finds correlations between $L_{b}$ and total observed
galaxy number, cluster velocity dispersion, and \xray luminosity, although not
to a high degree. A joint \xray and $K$-band analysis of 22 clusters finds weak 
correlation
between $L_{b}$ and cluster temperature \citep{edge91a}, which we now know is a 
good proxy for cluster virial mass. Subsequent studies
have confirmed these earlier findings \citep[e.g.][and references
therein]{fisher95,brough02,degrandi04}.  Our result also agrees qualitatively
with the previous investigations, but it is unique in the relatively large sample
size and its broad range in mass, the reliable cluster mass estimates (derived
from \xray temperature measurements), the physical quantities being evaluated
at fixed overdensities, and the use of $K$-band data, which is less sensitive
to on-going star formation.

What is less appreciated in previous studies is the BCG luminosity fraction,
i.e. the BCG-to-total luminosity ratio. There have been estimates of this 
fraction for some clusters (e.g. A2029, BCG+envelope, $\sim 23\%$ in $R$-band, 
\citealt{uson91}; A1651, BCG+halo, $\sim 36\%$ within 0.7 Mpc, in $I$-band, 
\citealt{gonzalez00}; three massive clusters at $0.8<z\lesssim1.0$, 
$10 - 16\%$ in $K$-band, \citealt{ellis04}).
These results show that indeed the BCG contribution to total cluster light is
large. With the large sample and the information on the total light from
cluster galaxies, we are in good shape to study the BCG luminosity fraction from
groups to massive clusters, which is plotted in 
Fig~\ref{fig:bcglf}. We recall that the total luminosity is obtained by summing
the contributions from the BCG and from other cluster galaxies, where the latter
component is obtained by integrating the LF of individual clusters.
We also note that the uncertainty in the ``total'' BCG light should not affect
the overall trend (except for the normalization), because, as shown in the previous section,
the magnitude differences between the $\mu_K=20$ mag/arcsec$^2$ isophotal
and deeper photometry do not depend on system mass.

As clearly shown in the figure,
the BCG light constitutes a very large fraction of total light at group mass 
scales ($40 - 50\%$), but becomes less important in the overall luminosity 
budget progressively, as we move to highest mass clusters ($\sim 5\%$). 
To be more quantitative, we find in paper II that the light from normal galaxies
in clusters grow as $M_{200}^{0.82\pm0.04}$; as long as the slope is greater
than that of the $L_b$--$M$ correlation, a decreasing BCG luminosity fraction is
a natural outcome.
Clusters appear to gain light faster than BCGs do, which may be the case when
a significant number of galaxies are accreted from either the field 
environments or less massive groups or clusters, while
at the same time the rate of BCG growth (e.g. via galactic cannibalism or 
mergers) remains rather modest. 
We explore this possibility as well as some BCG formation
and evolution scenarios in the next section.

\begin{inlinefigure}
   \ifthenelse{\equal{\figtype}{EPS}}{
   \begin{center}
   \epsfxsize=8.cm
   \begin{minipage}{\epsfxsize}\epsffile{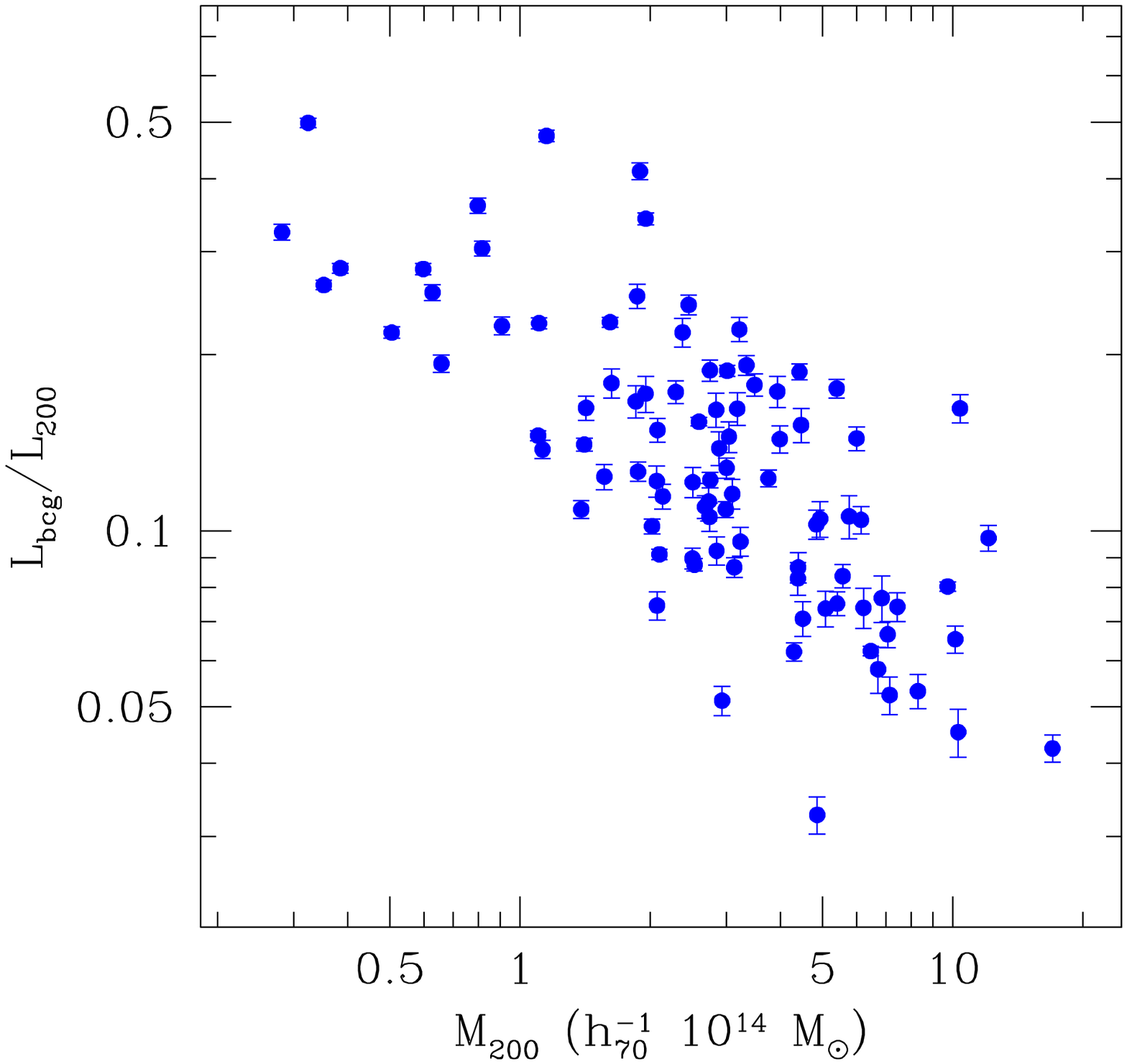}\end{minipage}
   \end{center}}
   {\myputfigure{f4.pdf}{0.0}{1.1}{-90}{-30}}
   \figcaption{\label{fig:bcglf}
	The BCG light fraction, defined as the ratio of BCG luminosity-to-total 
	cluster galaxy light (BCG and other galaxies).
	The comparatively low rate of galaxy mergers with the BCG as groups are
	merged hierarchically into more massive clusters leads to
	the decreasing BCG light fraction (see \S\ref{sec:tree}).
     }
\end{inlinefigure}

\subsection{Implications for BCG Formation and Evolution}
\label{sec:bcgform}

There are several proposals for the formation and evolution of the BCGs,
including ``galactic cannibalism'', a rapid merger among galaxies during the
collapse and virialization of low mass systems, and the cooling flows. 
In the first scenario,
orbital radii of galaxies decrease with time because of dynamical friction, and
eventually galaxies merge with the galaxy that first sinks to the bottom of the
gravitational potential. Therefore a central galaxy grows in mass and luminosity
as the cluster evolves \citep{ostriker75,richstone76,hausman78,malumuth84}. 
This picture has been 
criticized for requiring unreasonably short dynamical friction timescales, and 
overpredicts the luminosity gained through mergers over
cosmic time \citep{merritt83}. Instead, 
\citet{merritt84,merritt85} show that the dynamic friction time scale is very 
long compared to the Hubble time (due to the cluster mean field tidal stripping
of subhaloes), and the most likely way of
forming a giant central galaxy is by rapid merging at an epoch during the 
collapse and virialization of groups or low mass clusters. In the third theory, 
strong star formation activity takes place
at cluster centers because of the cooling of gas \citep[e.g.][]{cowie77,
fabian94a}. The absence of a large color gradient together with the red color of
BCGs seem to argue against this 
possibility, unless the stars formed are biased toward low mass \citep{edge91a}.
Furthermore, recent \xray observations suggest the ICM in ``cool-core'' clusters
never cools below $1-2$ keV \citep[e.g.][]{peterson01}, and so we can rule out
this scenario.

Between the other two scenarios, the rapid merging during low mass cluster 
collapse seems to
be more plausible. A high resolution $N$-body simulation which follows the
hierarchical build up of a Virgo-like cluster suggests the BCG does form in
a relatively short phase early in cluster evolution, through mergers with a few
massive galaxies \citep{dubinski98}. The alignment between the BCG 
semimajor-axis and the surrounding large-scale structure also seems to support
this idea, as the preferred direction for merger or accretion would be
along the large scale filaments \citep{west94}. Furthermore, there is direct
observational evidence of mergers involving the brightest members
taking place in intermediate and high redshift
clusters \citep{nipoti03,yamada02}.

Under this scenario, the natural formation sites for (progenitors of) BCGs are 
low mass systems, because the merger rate is a steeply declining function of the
velocity dispersion of a system \citep{merritt85}. The subsequent evolution of
the BCGs has to explain the existence of the $L_{b}$--$M$ correlation
(Fig~\ref{fig:bcglm}).  A simple picture is that BCGs form in groups and become
BCGs or other bright galaxies in more massive systems, as hierarchical
structure formation continues \citep{merritt85,edge91a}. After a
central dominant galaxy in the merged system is formed, cannibalism can take
place, but with a very small luminosity ingestion rate (e.g. $\sim 2 - 4\,L_*$
over a course of 10 Gyrs, \citealt{merritt85,lauer88,tremaine90}). Given that
$L_* \sim 10^{11} L_\odot$ (paper II), this amount seems to be able to account
for the mild increase of $L_{b}$ from group scales ($M_{200} \le 10^{14}
M_\odot$) to low or intermediate mass clusters ($M_{200} \le 5\times 10^{14}
M_\odot$), $\Delta L_b \sim 3 L_*$ (c.f. Fig~\ref{fig:bcglm}).  To produce the
more luminous BCGs in more massive clusters, other mechanisms may be needed.

Below we seek plausible scenarios based on the observations presented in the
previous section and in paper II. We first compare the luminosity functions
of high and low mass clusters and argue that BCGs in massive clusters are
likely built from luminous galaxies in progenitors of the host clusters
(\S\ref{sec:lfv}); then we show that the \lCDM model supports this scenario 
and leads to the observed decreasing BCG light fraction (\S\ref{sec:tree}).

Before we proceed, it is worth mentioning that we have adopted the view that the
BCGs in our sample represent a continuous evolution track. 
BCGs in the present-day high mass clusters may be evolved from low mass cluster
BCGs at higher redshifts, which may not be the same as those in the $z=0$
low mass clusters.  We have neglected this possible
difference in BCG properties in groups or clusters at different cosmic epochs;
we return to this issue in \S\ref{sec:bcgformdisc}.

\subsubsection{Constraints from Luminosity Function Variations}
\label{sec:lfv}

First we recall the conclusion drawn from 
Figs~\ref{fig:bcglm} \& \ref{fig:bcglf}: BCGs grow in luminosity, but at a less rapid rate compared to
their host clusters. Within the hierarchical structure formation paradigm, clusters
grow in mass and light by merging with other galaxy systems, and by accreting
isolated field galaxies. How can BCGs grow? We examine two possibilities:
(1) mergers with normal galaxies in the clusters, which, according to the 
calculations mentioned above, may provide a small amount of light for the BCG
growth. (2) mergers with luminous galaxies, which may likely be ex-BCGs of the
progenitors of the clusters under consideration. We discuss these options in
turn.

\begin{inlinefigure}
   \ifthenelse{\equal{\figtype}{EPS}}{
   \begin{center}
   \epsfxsize=8.cm
   \begin{minipage}{\epsfxsize}\epsffile{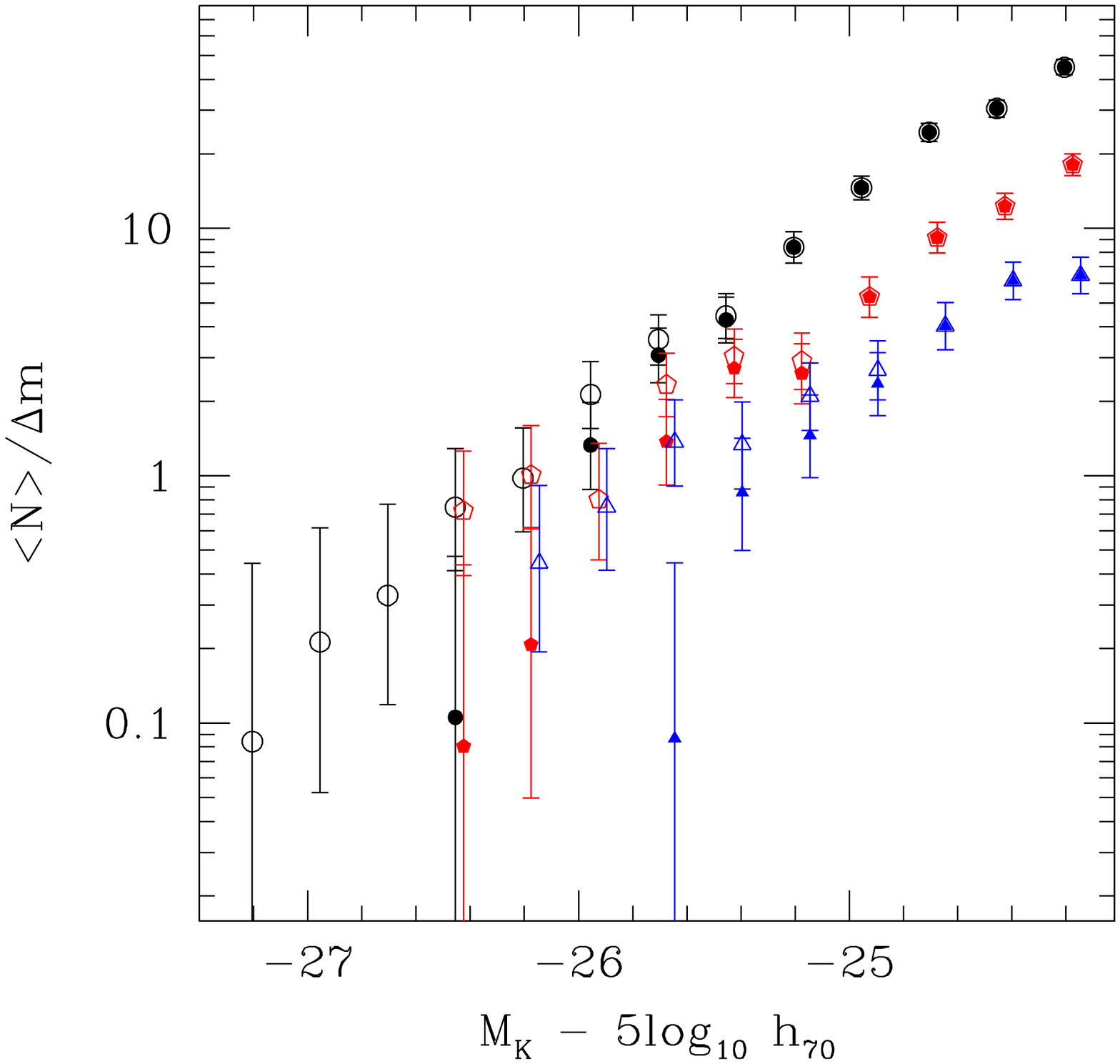}\end{minipage}
   \end{center}}
   {\myputfigure{f5.pdf}{0.0}{1.1}{-90}{-30}}
   \figcaption{\label{fig:ld}
	Galaxy luminosity distribution (LD: the mean number of galaxies per 
	cluster per magnitude, with binwidth $\Delta m = 0.25$ mag) in our 
	cluster sample. The circular,
	pentagonal, and triangular symbols represent average galaxy LDs from the
	most massive, the intermediate, and the least massive 25 clusters.
	The solid symbols represent the galaxy counts when BCGs are excluded,
	while hollow symbols shows the whole galaxy sample. Note that the most
	luminous galaxies are found only in the highest mass clusters. The
	comparison indicates that high mass cluster LD cannot be built from
	LDs of intermediate and low mass clusters without galaxy mergers.
     }
\end{inlinefigure}

In paper II (\S 5.1) we study the difference between the LFs of high and low
mass clusters by stacking the 25 most and least massive clusters (the mean 
masses are $M_{200,h} = 7.3\times 10^{14} M_\odot$ and $M_{200,l} = 1.1\times 
10^{14} M_\odot$, respectively). Two interesting points are found from the 
comparison: (1) the brightest galaxies appear only in the most massive clusters, and
(2) there is a marked deficit of $\sim L_*$ galaxies in high mass clusters,
compared to the low mass cluster LF. We note, however, these observations refer
to non-BCG galaxies, because no BCGs are included when we construct the LFs.
It is straightforward to calculate the amount of light contained in the
$\sim L_*$ galaxies that are ``missing'' in the high mass clusters.  We
integrate the difference of the low and high mass cluster LFs to obtain the 
luminosity density of the missing galaxy population, multiply it by the volume 
of a typical low mass cluster,
then by the mass ratio between high and low mass systems $M_h/M_l$. The amount 
of light is $\sim 2\times 10^{11} L_\odot \approx 2 L_*$, about 40\% of the
difference between BCG luminosities in high and low mass systems ($\sim 5L_*$). 
Therefore even if all the $\sim L_*$ galaxies that are missing in high mass
clusters were devoured by the BCG, there is still need for an additional
source to feed the central beast.

To explore the second possibility, we turn to the luminosity distribution (LD) 
in clusters of different mass scales. The LD is the mean number of galaxies per
cluster per magnitude as a function of magnitude \citep[e.g.][]{schechter76}; the difference 
between this and the luminosity function is that the LF contains information on 
spatial density of the galaxy populations.  Following similar
procedures outlined in \S 3.2 in paper II, we construct LDs for 3 mass scales;
in addition to the 25 most and least massive clusters, an intermediate mass
class of clusters (the mean mass being $M_{200,i} = 2.8\times 10^{14} M_\odot$) 
is also considered. We show in Fig~\ref{fig:ld} the resultant LDs.
In the figure different point styles correspond to different mass classes.
For the LD of each mass class we use two symbols to differentiate when the BCGs are included
(hollow points) or excluded (solid points). It is apparent that the luminosity 
of the brightest galaxies in each mass class increases with the mass of the 
clusters. Another interesting point is that in the intermediate or high
mass clusters, even non-BCG galaxies can be as bright as (or even brighter than)
the BCGs in the low mass clusters (e.g. solid circles/pentagons v.s. hollow
triangles). Finally, we note that there is on average $1.0^{+0.3}_{-0.2}$ galaxy
more luminous 
than $M_K = -25.3$ in a typical low mass cluster, while there are on average
$3.1^{+0.5}_{-0.3}$ and $2.0^{+0.4}_{-0.3}$ such galaxies in high and 
intermediate mass clusters, respectively.
However, if a typical high mass cluster is composed solely from either typical 
intermediate or low mass clusters (i.e. no significant amount of luminous
field/isolated
galaxies accreted, or no significant galaxy destruction during cluster merging),
we would expect there to be 5 or 6 galaxies brighter than 
$M_K = -25.3$, instead of 3, in the high mass clusters. Thus, galaxies appear to
be missing on the bright end in massive clusters.  

We can repeat the exercise 
for the amount of light expected.  The observed luminosity from galaxies 
brighter than $M_K = -25.3$ in a high mass cluster is $2.2\times 10^{12} 
L_\odot$, which is at most $2/3$ of the expectation if the light were coming
from all galaxies in the same luminosity range in intermediate or low mass 
clusters.  Therefore, from both galaxy counting and the luminosity budget, we conclude that
there are more than enough bright galaxies in lower mass 
systems to make up BCGs in the most massive clusters.  However, the massive 
cluster luminosity distributions are not simply the sum of many low mass 
luminosity distributions, because the massive clusters contain (1) fewer galaxies than would be expected in this case and (2) more luminous galaxies than any  in the low mass clusters.

\subsubsection{Merger Tree Considerations}
\label{sec:tree}

The data seem to support a scenario where, after forming in 
low mass groups, BCGs continue to grow primarily by merging with BCGs and $\sim L_*$ galaxies in 
other galaxy systems, as the host group/cluster grows hierarchically.
It is important to test whether the \lCDM cosmology supplies enough mergers for BCGs to
grow, and if cluster light grows faster than the BCG light. We utilize a merger 
tree algorithm developed by \citet{somerville99} to examine these issues. For a
halo of mass $M_0$ at $z=0$, we follow the accretion/merger history of its
most massive progenitor (MMP); more specifically, at each redshift step
($\Delta z \approx 0.05$), we record the number and mass of haloes which
merge with the MMP, the main ``trunk'' of the tree.
For a progenitor halo of mass $M_h$, we estimate its 
galaxy content by using the halo occupation number $N(M_h)$ determined directly 
from our cluster sample (paper II, \S 4.2). For simplicity we ignore the
difference between the definition of the halo mass (i.e. that suitable for 
theoretical mass functions and that used as the virial masses in our 
observations). We also ignore any time evolution in the halo occupation number
here (however, see \S\ref{sec:icllite} below).

\begin{inlinefigure}
   \ifthenelse{\equal{\figtype}{EPS}}{
   \begin{center}
   \epsfxsize=8.cm
   \begin{minipage}{\epsfxsize}\epsffile{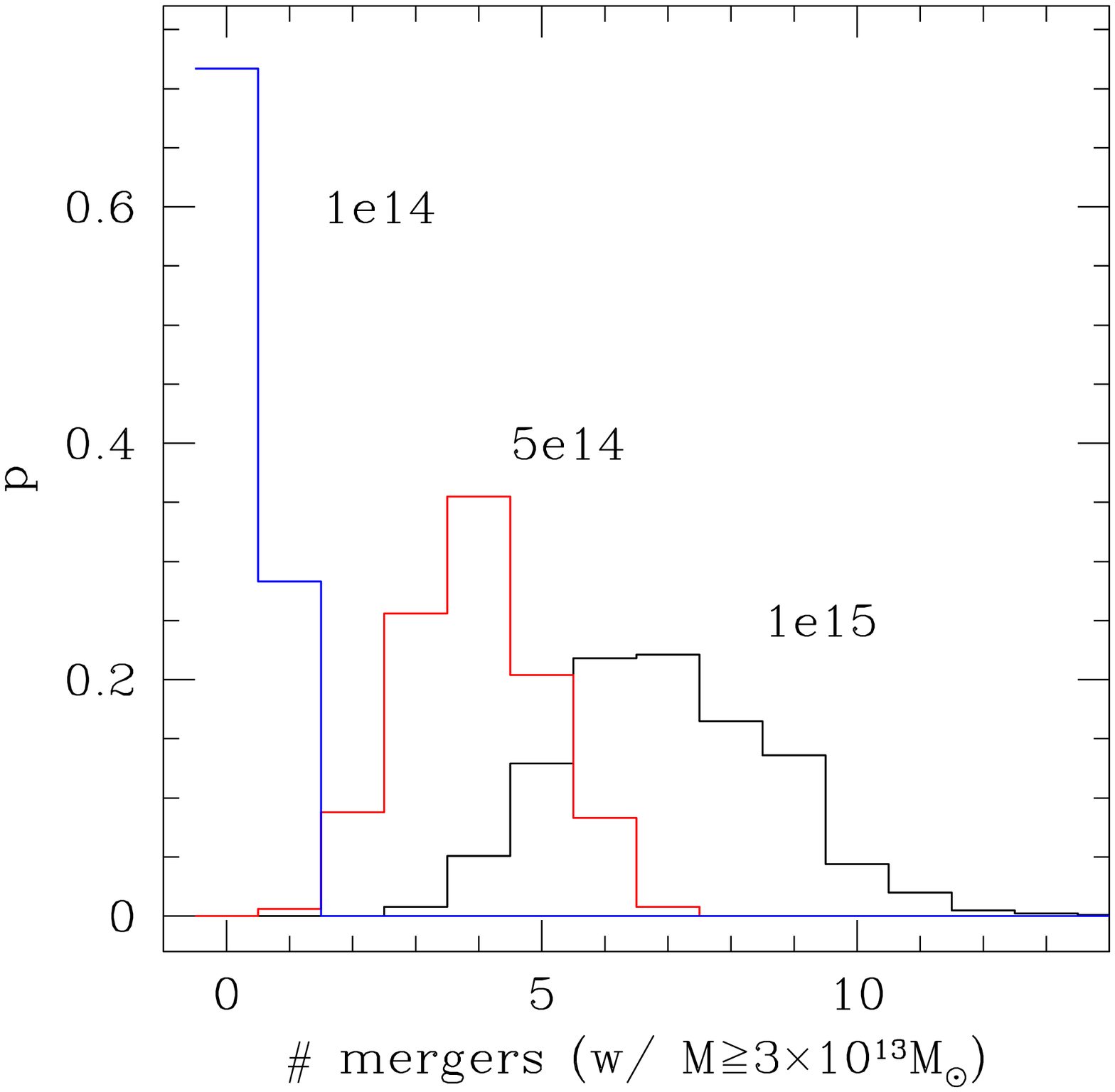}\end{minipage}
   \end{center}}
   {\myputfigure{f6.pdf}{0.0}{1.1}{-90}{-30}}
   \figcaption{\label{fig:merger}
	Probability distribution of the expected number of major mergers for 
	present day clusters of different masses. The results for each halo mass
	($10^{14}$, $5\times 10^{14}$ \& $10^{15} M_\odot$) are generated from
	1000 monte carlo realizations of merger processes, following the 
	extended Press-Schechter formalism outlined in \citet{somerville99}.
	These distributions provide an indication of how many BCGs are delivered
	from lower mass scale groups and clusters in the formation of systems
	with these three masses.
     }
\end{inlinefigure}

We first consider the BCG growth. We estimate the numbers of massive groups 
($M_h\ge 3\times 10^{13}M_\odot$) that are expected to merge onto the MMP of a 
present day $M_0=10^{14}M_\odot$, $M_0=5\times 10^{14}M_\odot$ \& 
$M_0=10^{15}M_\odot$ system. 
With the power spectrum normalization $\sigma_8$ adopted from the the 
{\it WMAP} result \citep{bennett03} we
find the numbers to be $0.3^{+0.5}_{-0.3}$, $3.9\pm1.1$ \& $7.2\pm1.8$, 
respectively (Fig~\ref{fig:merger}). 
Assuming there is at least one galaxy as bright as the observed 
BCGs in our lowest mass clusters ($3-4L_*$) for each of the
groups merged, and that these galaxies do not get disrupted 
during infall, there appears to be more than enough light to account for the
luminosity differences between
BCGs in each of the mass classes (c.f. Fig~\ref{fig:bcglm}). Furthermore, if the
merger efficiency between the brightest galaxies is such that one ex-BCG from 
every $2-3$ merged groups merges with the BCG of the main halo, the observed 
difference in the number of bright galaxies can be accounted for.

Next let us consider the number of (normal) galaxies to be accreted/merged with
the MMP.  We assume that a halo whose mass is in the range $10^{11} \le M_h
\lesssim 10^{13} M_\odot$ contains only one galaxy, while more massive haloes
follow the observed $N(M_h)$ relation.
We note that the observed $N(M_h)$ relation is derived for galaxies brighter
than $M_K=-21$; from the observed galaxy luminosity function \citep{kochanek01}
and the theoretical halo mass function \citep{sheth99} we estimate the mass
of such galaxies to be $\sim 10^{11} M_\odot$.  For $M_0=10^{14} M_\odot$,
$5\times 10^{14}M_\odot$, \& $10^{15} M_\odot$, we find that there are
$57\pm4$, $205\pm11$ \& $373\pm16$ galaxies to be accreted, respectively
(among these are $36\pm6$, $51\pm5$ \& $54\pm5$ isolated galaxies).
We expect these galaxies to lose mass as they enter denser environments.
Although it is possible to estimate the mass loss given a model for halo
structure \citep[e.g.][see \S5.1 in paper II]{klypin99}, it is not clear how the
light in the galaxies will be affected.
Under the assumption that the majority of these galaxies are not destroyed by
tidal forces, we expect that clusters grow
in their galaxy content by mainly merging with other groups/clusters, and the
number of these galaxies outnumbers the ex-BCGs.
We show in paper II (\S 4.2) that for
non-BCG galaxies in clusters, light from galaxies roughly traces galaxy number
\citep[see also][]{rines04}. It thus appears that the increase in galaxy number
due to merging with other clusters, groups and galaxies infalling from the
field can account for the decreasing BCG light fraction.

It remains to be seen if the merger processes are short enough that the 
brightest
galaxies can actually merge. This process is considered by \citet{tremaine90}.
Consider a merger between two clusters, where both have
a giant galaxy at the center. It is possible that the two giants will end up
orbiting each other, and eventually merge in a time scale of about $1/5$ of the
Hubble time \citep{tremaine90}. The ``dumbbell'' galaxies observed in some clusters, usually composed of two D or cD galaxies, may be in the process of merging 
\citep[e.g.][]{quintana96}. Although the merger time scale may be short enough, 
definitive statements require more sophisticated numerical experiments.

\subsection{Second and Third Ranked Galaxies}
\label{sec:m2m3}

If clusters form hierarchically, the galaxies on the bright end of the LF may be
themselves BCGs in progenitors that merge to form the current cluster. In the
picture outlined in the previous section where the BCGs most likely grow in
luminosity by merging with other bright galaxies, we can examine properties
(e.g. luminosity, kinematics, etc) of the second or third ranked galaxies (G2 
\& G3), in hopes that they may provide some constraints on the scenario.  

Using redshift information from the NED, we have secure membership assignments 
for 78 clusters down to G3 (82 clusters have confirmed G2). We treat
these galaxies the same way as we do the BCGs (photometry, $k$-correction,
etc). The luminosity ($L_{G2}$, $L_{G3}$) and the luminosity fraction
($L_{G2}/L_{200}$, $L_{G3}/L_{200}$) are shown in Fig~\ref{fig:l2l3}.
We see that these galaxies show trends similar to the BCGs: the luminosity
for both G2 and G3, although increasing mildly with cluster mass, 
becomes less important compared to total light in all galaxies in more massive 
systems. It is interesting to see that on average the G2s in massive clusters
($M_{200}\ge5\times10^{14}M_\odot$) are about as luminous as the BCGs in lower
mass clusters ($M_{200}<10^{14}M_\odot$).
In the lowest mass groups in our sample, the sum of the luminosities from the 
brightest three galaxies accounts for nearly the total light, while their sum 
is about 10\% of total light in the most massive clusters. 
The similar correlations between the galaxy luminosity and the cluster bulk
properties (mass and luminosity)
suggest these brightest galaxies may share a similar formation history.

\begin{inlinefigure}
   \ifthenelse{\equal{\figtype}{EPS}}{
   \begin{center}
   \epsfxsize=8.cm
   \begin{minipage}{\epsfxsize}\epsffile{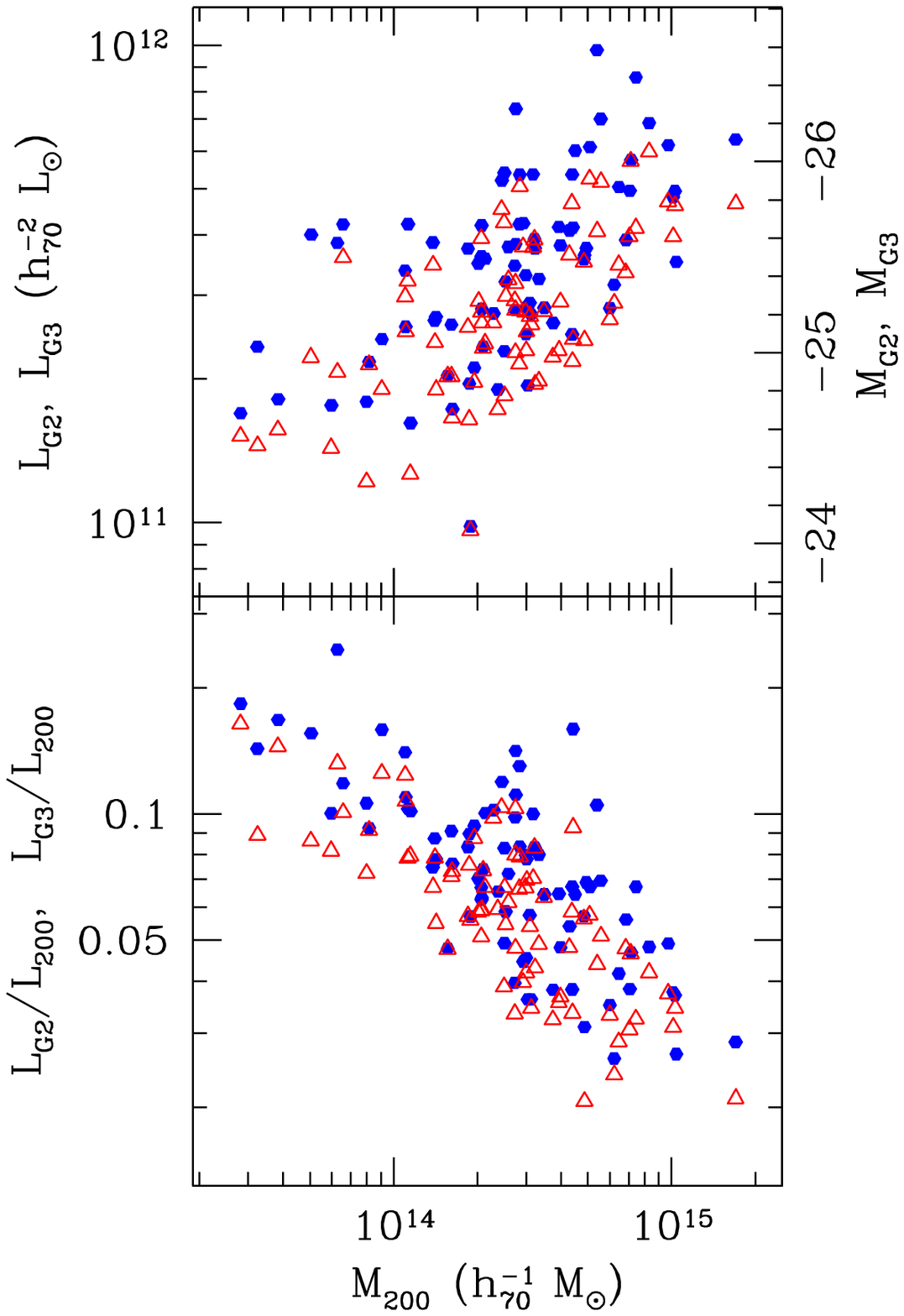}\end{minipage}
   \end{center}}
   {\myputfigure{f7.pdf}{0.0}{1.1}{-90}{-30}}
   \figcaption{\label{fig:l2l3}
	Correlations between the luminosity of the second (G2, solid points) and
	the third (G3, hollow triangles) ranked galaxies and cluster properties.
	Upper panel: Both G2 \& G3 become more luminous as cluster mass 
	increases. Lower panel: G2 \& G3 luminosity-to-total luminosity 
	fraction.  On average the second ranked galaxies are 0.24 mag
	brighter than the third ranked galaxies (based on 78 clusters where
	information is available) and 0.66~mag fainter than the BCGs.
     }
\end{inlinefigure}

The average magnitude difference between the BCGs and G2s is
0.66 mag, with a dispersion of 0.48 mag; the same quantities for the BCGs and 
the G3s are 0.89 mag \& 0.45 mag, respectively. The average
magnitude difference between G2 and G3 is 0.24 mag, with a dispersion of 0.22 mag. We also note
there is no trend between the luminosity difference and the cluster mass.
Our results can be compared to that of \citet{schneider83b}; the difference
between the mean absolute magnitudes of BCGs and G2s is 0.83 mag (in $r$ band),
and that between the BCGs and G3s is 1.32 mag. Notice that their photometry is 
measured within a fixed metric radius of $13.7h_{70}^{-1}$ kpc.

We also examine the projected distribution of the G2s with 
respect to the BCGs, as a function of cluster mass. We find no apparent evidence
that suggests the G2s in more massive clusters lie closer to
the BCGs. As for the velocity differences between the two brightest galaxies,
a Kolmogorov-Smirnov test indicates a weak possibility that the velocity 
differences (scaled by the cluster mean velocity dispersion) in clusters
more massive than $M_{200}=2.8\times 10^{14} M_\odot$ is different from 
those found in lower mass clusters (a 9\% chance that the two distributions are 
the same).  These findings do not provide strong indications of a kinematic relationship between the BCGs and G2s; we suggest that a larger cluster sample with a targeted spectroscopic study would be an interesting way to further probe for a relationship.

\section{Constraints on Intracluster Light}
\label{sec:icl}

It has long been suggested that dynamical processes will alter or transform
properties of galaxies that orbit in clusters 
\citep[e.g.][and references therein]{gunn72,merritt84,moore98b,treu03}.
Mergers, galaxy harassment, tidal
truncation, and ram pressure stripping may all liberate galactic material
(dark matter, stars, gas, etc) into the intergalactic space 
\citep[e.g.][]{malumuth84,gnedin03b, napolitano03,sommerlarson04,murante04,willman04}. 
Observations have provided ample evidence to support this inference. Intracluster
stellar populations are detected in various forms, as reviewed in \S\ref{sec:intro} (see also \citealt{arnaboldi03b,feldmeier03c} for more complete account).
These may well be the source of the ICL, as well as the diffuse
haloes/envelopes of the cD galaxies. We emphasize here that no distinction
between ICL and cD envelope is made in this paper.

There must be another component in clusters that accounts for the light lost from the galaxies.
This component is bound to the cluster, but may or may not be distributed like
the galaxies. We identify this component as the ICL. In general, we expect ICL
to become more and more important in the overall light budget in massive clusters.
However, the detailed behavior would of course depend on the complicated 
dynamical history of clusters, as well as the galaxy formation efficiency across
the mass spectrum of clusters. In \S\ref{sec:icllite} we propose simple models
to estimate the ICL light fraction as a function of cluster mass, and compare
these predictions with available observational constraints.

In addition to confirming this dynamical view of galaxy--cluster interactions,
studies of the ICL may provide a more accurate determination of the cluster
baryon fraction and cold fraction (defined as the ratio between stellar mass
and total baryon mass in clusters), which in turn should help clarify the star
formation efficiency in clusters (\S\ref{sec:iclbf}).
Finally, the ICL may serve as an important source of the enrichment
of the intracluster medium (ICM). The ICM is known for its high metal abundance
($\sim 0.3 Z_\odot$ for iron, \citealt{degrandi04} and references therein). 
Apparently the metal comes from stars, i.e. via supernova explosions. The large 
amount of metal present in the ICM can not be accounted for with standard
chemical yields or metal transport mechanisms such as galactic 
winds \citep[e.g.][]{portinari04}; confined by the potential wells of individual
galaxies, the efficiency of the enrichment processes is not high.
On the contrary, supernovae orbiting in the 
intracluster space surrounded by the ICM or supernovae in low mass, pregalactic 
structures, may be much more efficient agents of enrichment. In 
\S\ref{sec:enrich} we use the predicted amount of ICL to reevaluate the 
enrichment problem in the ICM (see also \citealt{zaritsky04}).

\begin{table*}[htb]
\begin{center}
\caption{Some Observational Constraints on the ICL Fraction}
\begin{tabular}{lcccccc}
\tableline \tableline
Name & Virial Mass 			& Band & Depth & Extent & $f_{i,2D}$ & Reference \\
     & ($10^{14} M_\odot$)  &  /Method  & (mag/arcsec$^2$)  &    &       &    \\
\tableline 
A2390 &  14.9      & r   & 26.7 & 204 kpc ($0.09\, r_{200}$) & $0.02\pm 0.06$\tablenotemark{\dagger} & 1 \\
A2029 & 11.2   & R   & 26.0 & $360-640$ kpc ($0.18-0.32\, r_{200}$) & 0.1 & 2 \\
A1656 & 9.7     & R  & 27.7  & $8'$ ($0.12\, r_{200}$) &  $0.5$     & 3 \\
A1689 & 9.6     & R  & 28.2  & 143 kpc ($0.07\, r_{200}$) & $0.3$ & 4 \\
A1914 & 9.4     & V  & 26.0  & 1.6 Mpc ($0.84\, r_{200}$) & $0.28$ & 5 \\
A1413 & 6.4     & V  & 26.5  & 180 kpc ($0.10\, r_{200}$) & $0.13$ & 6 \\
A1651 & 6.2     & I  & 29.5  & 714 kpc ($0.42\, r_{200}$) & $0.38$\tablenotemark{\sharp} & 7 \\
Cl0024+1652 & 3.4 & V & N/A & 153 kpc ($0.12\, r_{200}$) & $0.15\pm 0.03$ & 8 \\
A2670 & 2.7      & R  & 29.0  & $3.5'$ ($0.23\, r_{200}$) & 0.3 & 9 \\
Virgo & 1.5    & ICPNe\tablenotemark{\ddagger} & 27.4   & 0.196 deg$^2$ & 0.10\tablenotemark{\dagger} & 10,11 \\
      &        & ICPNe\tablenotemark{\ddagger} & 27.0   & 0.89 deg$^2$ & $0.16\pm0.03$\tablenotemark{\dagger} & 12 \\
      &         & I/ICRGB\tablenotemark{\S} & 27.9 & 4.73 arcmin$^2$ & $0.1-0.2$\tablenotemark{\dagger} & 11 \\
HCG90 & 0.19    & V  & 24.5  & 64 arcmin$^2$ & $0.45\pm0.05$ & 12 \\
M81 group &  0.012\tablenotemark{\flat} & ICPNe\tablenotemark{\ddagger}  & N/A & 1.44 deg$^2$ & 0.013\tablenotemark{\dagger} & 13 \\
Leo I group & N/A & ICPNe\tablenotemark{\ddagger}  & 24.7  & 0.26 deg$^2$ & $<0.016$\tablenotemark{\dagger} & 14 \\
\tableline
\end{tabular}
\tablecomments{ 
    $h_{70}=1$ is assumed in calculating the mass $M_{200}$ and size. 
    References: (1) \citet{vilchez94}; (2) \citet{uson91};
    (3) \citet{bernstein95}; (4) \citet{tyson95}; (5) \citet{feldmeier04};
    (6) \citet{feldmeier02};
    (7) \citet{gonzalez00}; (8) \citet{tyson98}; (9) \citet{scheick94};
    (10) \citet{okamura02} (11) \citet{arnaboldi03} (12) \citet{feldmeier04b};
    (13) \citet{durrell02}; (14) \citet{pwhite03};
    (15) \citet{feldmeier03b}; (16) \citet{castro03}.
    $^\dagger$: excluding BCG envelope. $^\sharp$: including BCG light.
    $^\ddagger$: derived from intracluster 
    planetary nebula (ICPN) abundance. $^\S$: derived from intracluster RGB star
    (ICRGB) abundance. $^\flat$: \citet{karachentsev02}.
  }
\end{center}
\vskip-35pt
\end{table*}

\subsection{Estimating the Intracluster Light}
\label{sec:icllite}

\begin{inlinefigure}
   \ifthenelse{\equal{\figtype}{EPS}}{
   \begin{center}
   \epsfxsize=8.cm
   \begin{minipage}{\epsfxsize}\epsffile{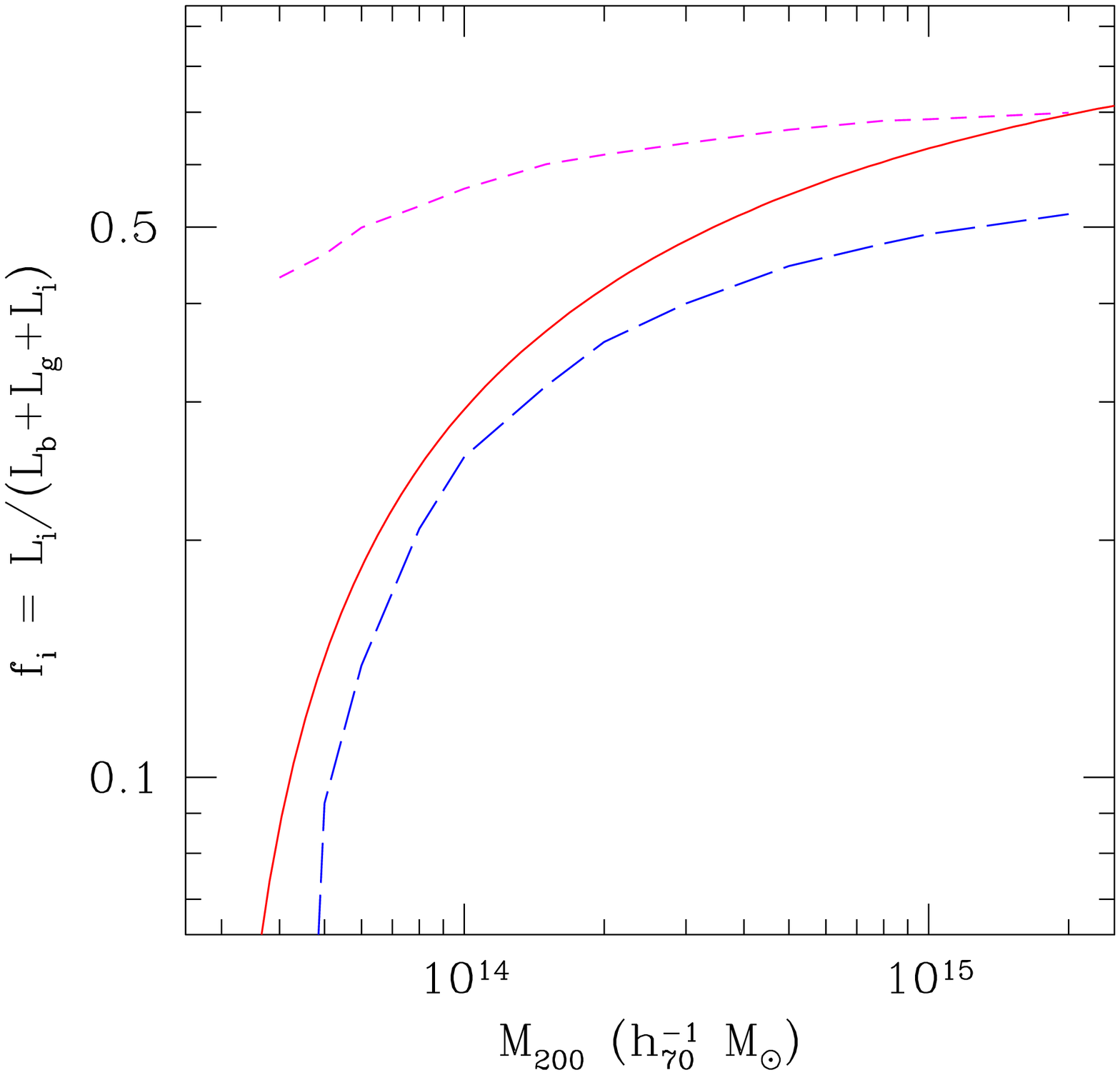}\end{minipage}
   \end{center}}
   {\myputfigure{f8.pdf}{0.0}{1.1}{-90}{-30}}
   \figcaption{\label{fig:icl}
	Estimated ICL fraction. The solid curve corresponds to the prediction of
        Eqn~\ref{eq:icl}, with $\sim 1\%$ of light in ICL in lowest mass groups
        ($M=3\times 10^{13} M_\odot$). The two dashed lines are the predictions
        from a merger tree model (Eqn~\ref{eq:icl2}), with two models for 
        redshift evolution of the light--mass relation (long-dashed: 
	no-evolution, $\gamma=0$; short-dashed: $\gamma=1$, evolution consistent
	with numerical simulations).
     }
\end{inlinefigure}

We decompose the total light in a cluster into three components:
\[
L_{tot} = L_b + L_g + L_i,
\]
where the terms on the right hand side denote the light of the BCG, the other
galaxies, and the ICL, respectively. In general, we expect all these components 
to be 
functions of cluster mass. Next suppose there is a system (of mass $M_\ell$) that 
serves as a building block of systems more massive than $M_\ell$. We denote
\[
L_{tot,\ell} = L_{b,\ell} + L_{g,\ell} + L_{i,\ell}
\]
for this system.  If clusters grow in mass and luminosity mainly by mergers
between such systems (i.e. accretion of field galaxy populations is not
important),
then we can write $L_{tot}(M) = (M/M_\ell) L_{tot,\ell}$; in this case 
the ICL light for a system of mass $M$ is simply
\begin{equation}
\label{eq:icl}
L_i(M) = \left(\frac{M}{M_\ell}\right) L_{tot,\ell} - L_b(M) - L_g(M).
\end{equation}

If we set the mass scale $M_\ell$ to be that of the least massive group in our 
sample
($\sim 3\times 10^{13} M_\odot$), using our observed behavior for $L_b(M)$ and 
$L_g(M)$, we would then be able to predict the amount of ICL as a 
function of cluster mass, with the value for ICL in groups of mass $M_\ell$,
$L_{i,\ell}$, being a free parameter.  This approach is effectively saying that 
the lower galaxy luminosity per unit cluster mass in massive clusters (recall
that the slope of the $L$--$M$ relation is less than unity) 
is simply because stars have been stripped off galaxies during the hierarchical 
merging that produced the high mass clusters from the low mass clusters.

In Fig~\ref{fig:icl} we show the predicted ICL light fraction $f_i \equiv 
L_i/L_{tot}$ of this toy model (solid line), assuming that $q \equiv 
L_{i,\ell}/(L_{b,\ell}+ L_{g,\ell}) = 0.01$, that is, an ICL that is 1\% of total light 
in galaxies for groups of mass $M_\ell$. The predicted ICL fraction is high. For clusters
more massive than $10^{14} M_\odot$, $f_i > 0.3$; at $10^{15} M_\odot$, about
2/3 of total light is in the intracluster space! This estimate does not change much if we
choose different values for $q$; for $q=0.05$ or $q=0.10$ the ICL fraction is
similar to the curve of $q=0.01$ for systems more massive than $10^{14} 
M_\odot$.

Effectively, Eqn~\ref{eq:icl} assumes that the stellar mass fraction is constant
in clusters of different mass, which may not be the case. Furthermore, the toy
model neglects the accretion of isolated galaxies, which
may be important (\S\ref{sec:tree}). We therefore consider a
second model which utilizes the merger tree algorithm used in \S\ref{sec:tree}.
The basic idea is to estimate the total light content of a cluster by counting
luminosities contained in all the galactic systems that fall into the cluster,
and compare it with the observed light in cluster galaxies. The difference would
then be the ICL. To this end, we need a light--mass relation for galactic 
systems ranging from single galaxies to clusters. For systems more massive than
$10^{13}M_\odot$, we employ the observed $L$--$M$ relation (\S4.1 in paper II); for
less massive systems, which we assume to be single galaxies, we estimate their
$L$--$M$ relation by matching the number densities indicated by the observed
$K$-band galaxy luminosity function \citep{kochanek01} and that predicted by the
theoretical halo mass function (modulated by the halo occupation number; 
\citealt{kravtsov04}).

By summing over the light of each galactic system that has merged directly onto
the main trunk of the tree (the MMP), we can estimate the total light as a 
function of the cluster mass. To be more specific, for a present-day cluster of 
mass $M_0$, we calculate
\begin{equation}
\label{eq:icl2}
L_{tot}(M_0) = \sum L_h(M_h)\, (1+z_h)^\gamma,
\end{equation}
where $M_h$, $L_h$ and $z_h$ correspond to the mass, the luminosity of the
merged haloes, and the epoch of the merger, respectively. The redshift 
dependence is included to account for any evolution in the $L$--$M$ relation or
halo occupation number \citep[][\S5.3 in paper II]{kravtsov04}. We consider two 
possibilities: $\gamma=0$ \& $\gamma=1$. The former is the no-evolution case,
while the latter is consistent with the halo occupation number evolution as 
suggested by numerical simulations \citep{kravtsov04}.

The predictions of this second model are shown in Fig~\ref{fig:icl} (averaged
over 100 merger tree realizations): 
The long-dashed line refers to the $\gamma=0$ case, while the short-dashed line 
shows the $\gamma=1$ results.  Interestingly the model results for the two 
values of
$\gamma$ enclose the predictions of the first model. We note that the 
second model does not account for any ICL contained in the galactic systems that
merge with the main halo; the predictions thus are lower limits on the ICL 
light fraction.

To compare these predictions with estimates derived
from direct observations, we compile in Table 2 several published values. The columns give the name of the cluster, the virial mass estimated
from published \xray measurements, the band used, the physical or angular extent
of the observation, the measured ICL fraction, and the original reference. 
Some remarks may help the reader.
First of all, these studies use different bands, have different depth, 
and survey different fractions of the clusters. We caution that the ICL fraction
may depend on the waveband used \citep{vilchez94}.  Our estimate is given in 
$K$-band for galaxies brighter than $M_K=-21$ mag (recall that we integrate the 
LF to $M_{min} = -21$ mag, \S\ref{sec:sample}).  For
the nearby clusters or groups, it is possible to estimate the ICL fraction from
tracers such as intracluster planetary nebulae (ICPNe) or RGB stars. However,
these techniques look for patches in the cluster/group fields and we only
list the total area of the observations, but not the fraction of the virial 
radius surveyed. Finally, strictly speaking, the ICL fraction
listed in the Table is the projected value, while the prediction of our model is
that contained in the 3D cluster region. To compare the values a model
for the radial distribution of ICL is needed.

In addition to these observations, a recent study stacking a large number of
clusters suggests the ICL contributes about $15-20\%$ of cluster optical light
within 500 kpc \citep{zibetti04}. A recent measurement of BCG+ICL profile for
a sample of 24 clusters is presented in \citet{gonzalez04}.

Finally, we compare these simple models with the findings from numerical 
studies. The modeling by
\citet{napolitano03} uses an $N$-body simulation of a Virgo-like cluster 
formed in a hierarchical fashion to study the distribution and correlation
properties of tracers of ICL. They estimate that outside the central region,
the ICL fraction may be $30-50\%$. A big step forward toward theoretical
modeling of the ICL is realized in the recent cosmological hydrodynamic 
simulations \citep{murante04,sommerlarson04,willman04}.
In particular, the former study finds that the ICL
fraction increases with cluster mass, in broad agreement with our results.
At $M_{200}=10^{14} M_\odot$ their model predicts $f_i \approx 0.25$, while at
$M_{200}=7\times 10^{14} M_\odot$ the ICL fraction is about 0.45; although it is
encouraging that both these values are very close to the predictions of our
second model with $\gamma=0$, a comparison between our local clusters and
clusters at higher redshifts suggests a nonnegligible evolution in the halo
occupation number (see \S 5.3 of paper II). Further investigations on the halo
occupation number evolution, as well as higher resolution numerical studies,
are needed to resolve this discrepancy.

The two models presented here, as well as other numerical studies, suggest that
the ICL can be very important in the overall cluster light budget; we discuss 
some of the implications next.

\subsection{ICL Contribution to Cluster Baryon Fraction}
\label{sec:iclbf}

An immediate implication of these estimates is that the baryon fraction and cold
fraction derived from galactic light need revision. Because of the relatively small contribution
of stars in galaxies to the baryon budget (\S4.1 in paper I), a roughly 100\% 
increase in stellar mass would not change the baryon 
fraction or the derived \Om constraints much.  Within our sample of 93
clusters, we have measured ICM gas mass for 35 (which we denote as the MME 
subsample, \citealt{mohr99}). Based on these clusters, the
mean baryon fraction is 0.1441, while for ``hot'' clusters (which we choose to
have $kT_X \ge 3.7$ keV, see \S3 in paper I), the mean is 0.1512. Using the 
value for baryon density from {\it WMAP} \citep{bennett03}, these correspond to 
$\Omega_M = 0.32$ \& 0.30, respectively. Now, from Fig~\ref{fig:icl} we can 
roughly estimate that the total
light (including the ICL) would be $\sim 2 (L_b + L_g)$. This assumption leads
to $\Omega_M = 0.29$ \& 0.28, for the MME subsample and the hot clusters within
it, respectively. Both these values are in good agreement with the {\it WMAP} 
estimate of $\Omega_M=0.27 \pm 0.04$. 

The presence of ICL has a more significant effect on the cold fraction; 
for the MME subsample and the hot clusters within it, the mean values are 0.150
\& 0.163, respectively. Compared to the mean values when no ICL is present
(c.f. \S4.2 in paper I), these correspond to a 40\% increase. Interestingly,
a recent cosmological hydrodynamical simulation \citep{borgani04} finds that 
roughly 20\% of baryons are cold for clusters hotter than 3 keV (roughly 
$M_{200} \sim 2\times10^{14} M_\odot$). The current implementation of 
feedback/heating mechanisms may not be far from producing the cold fraction in
real clusters.

\subsection{Enrichment of Intracluster Medium} 
\label{sec:enrich}

It is well established that the ICM contains an enormous amount of metal
\citep[e.g.][]{arnaud92,loewenstein96,finoguenov00,degrandi01,
baumgartner03}. 
Within standard models for star formation in
galaxies and the processes that transport metal to the intracluster space,
a standard stellar initial mass function (IMF) with typical supernova 
yields can not account for the metal production 
\citep[e.g.][hereafter P04]{portinari04}.
Non-standard scenarios such as a time-varying IMF, IMFs that differ in different
environments, or different enrichment agents (e.g. hypernovae from
population III stars) have been proposed as alternatives 
\citep[e.g.][P04]{finoguenov03,loewenstein01}.

Here we investigate to what degree the ICL 
can help account for the extraordinary metal 
production in clusters. We will focus on the iron abundance. Following the same
procedures outlined in \S 4.3 in paper I (with the latest ICM iron abundance 
from a sample of 22 clusters observed with {\it BeppoSAX}, $Z_{ICM,Fe} = 0.34
Z_{\odot,Fe}$, \footnote{this is the value based on the revised solar 
photospheric abundance \citep{grevesse99}, which implies iron mass fraction 
$X_{\odot,Fe} =0.0012$.}
\citealt{degrandi04}, hereafter DG04), we estimate the iron yield \citep{arnaud92}
$y_{Fe} = (M_{Fe,ICM}+M_{Fe,star})/M_{star}$
and the iron mass fraction (the iron-to-total mass ratio) for the MME subsample.
Two important quantities involved in the calculations of iron mass in stars are
the mean stellar mass-to-light ratio and the metallicity. We use the observed 
typical stellar mass-to-light ratio and metallicity for early and late type
galaxies, weighted by the relative abundance of the two types in clusters as a 
function of cluster mass, to account for possible variations of these quantities
with respect to cluster mass (see Appendix of paper I for more details).
In Fig~\ref{fig:iron} (upper panel) we show the iron yields obtained
without ICL (shown as hollow points). As we find in paper I, the iron 
yield is very high ($3-9 Z_{\odot,Fe}$) compared to the solar vicinity ($0.9-0.95 Z_{\odot,Fe}$, P04) 
and is an increasing function of cluster mass, while the iron mass fraction is 
roughly constant with respect to cluster mass.

The inclusion of the ICL reduces the iron yield. With the amount of ICL
predicted by our first model (Eqn~\ref{eq:icl}, c.f. Fig~\ref{fig:icl}), the
corresponding iron yield falls to about $3 Z_{\odot,Fe}$ (solid points in 
the
upper panel). We note, however, in this case the iron mass fraction 
(total iron mass over total binding mass; not shown in this figure) becomes a
weakly increasing function of cluster mass.

\begin{inlinefigure}
   \ifthenelse{\equal{\figtype}{EPS}}{
   \begin{center}
   \epsfxsize=8.cm
   \begin{minipage}{\epsfxsize}\epsffile{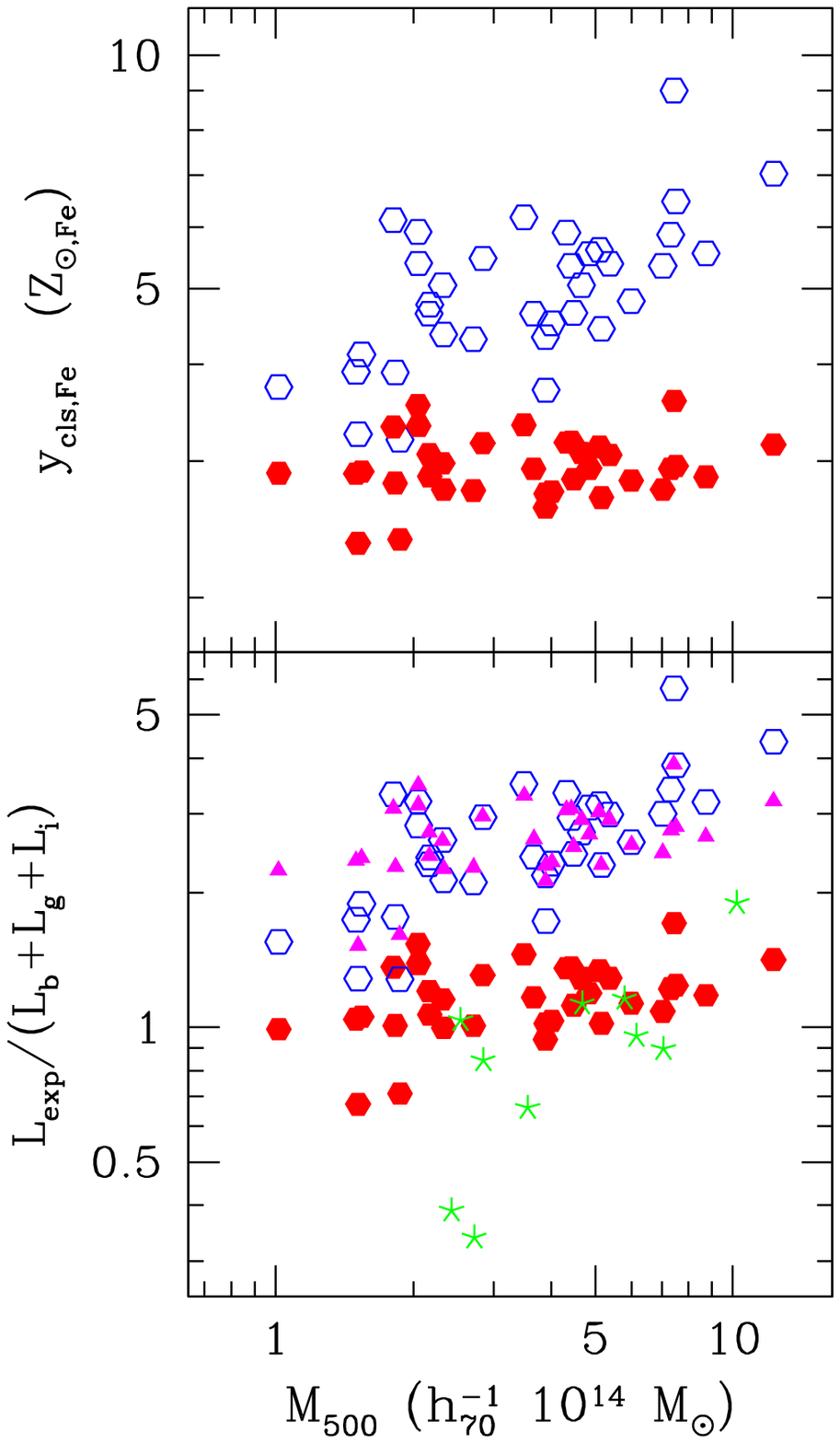}\end{minipage}
   \end{center}}
   {\myputfigure{f9.pdf}{2.3}{2.0}{-160}{-70}}
   \figcaption{\label{fig:iron}
	Upper panel: the iron yields in the MME subsample.
        The hollow points show the results derived without any ICL, while the
        solid symbols are the yields when an ICL given by Eqn~\ref{eq:icl} is
        included. The ICL reduces the yields and makes them roughly
        constant for clusters of all masses.
        Lower panel: the ratio between the total light required to produce 
	the observed iron (in stars and ICM) and the total observed cluster 
	light. The solid hexagons and triangles show the cases for Salpeter \& 
	Kroupa IMFs, respectively, and in these cases the ICL contribution to 
	the total light is included. The hollow points show the Salpeter case 
	with no ICL included. Note that in the Salpeter IMF model, there is 
	enough stellar light  (when ICL is included) to explain the enormous 
	iron reservoir.  The stellar symbols show the ratio of the luminosity 
	required to 
	produce the ``iron excess'' in cool core clusters compared to ICL+BCG 
	luminosity.  In most cases the ICL is crucial in producing the excess 
	iron component.
	Note the ordinate is $M_{500} \approx 0.72 M_{200}$, the 
	mass enclosed by the radius $r_{500}$ (see \S\ref{sec:sample}).
     }
\end{inlinefigure}

To further quantify the possible role of ICL in the ICM enrichment, we compare
the predicted iron mass-to-total light (ICL and galaxies) ratio (IMLR, 
\citealt{renzini93}) with the
single stellar population (SSP) models of P04. Specifically, 
given an initial mass function, a SSP model predicts the evolution of stellar
mass and luminosity, as well as its iron production. One can thus compute the
IMLR at any epoch. For a stellar population of 10 Gyr old, we calculate the
total light required for a cluster IMLR to be consistent with the SSP
prediction. We show the ratio between the required luminosity ($L_{exp}$) and our observed
luminosity ($L_{tot}\equiv L_b+L_g+L_i$) for each cluster in
Fig~\ref{fig:iron} (upper panel), with ICL given by Eqn~\ref{eq:icl}. The solid 
hexagons and triangles show the cases
for the \citet{salpeter55} and the \citet{kroupa02} IMF, respectively. We note
that the latter IMF describes the solar neighborhood better than the former 
(P04).  For comparison, we also
show the case for the Salpeter IMF when no ICL is included (hollow symbols).

From this figure we draw the following conclusions:
(1) for the Salpeter case, the observed light plus the ICL
is slightly lower than that required to produce the observed iron (the mean
is $\overline{L_{exp}/L_{tot}} = 1.17\pm0.04$; no theoretical uncertainty is included), 
but there is a severe shortage of stars and luminosity when a Kroupa IMF is used to produce
the observed iron (the mean is $2.66\pm0.08$); (2) when no
ICL is present, even with the Salpeter IMF, the required total light is about
2.7 times the observed light from all galaxies, and with the Kroupa IMF, the 
situation worsens to a factor of 6 more light required; (3) the iron production efficiency
becomes more uniform across different mass scales when the ICL is included, as can be inferred from the shallow slope ($0.14\pm0.05$ for both IMFs) of the $L_{exp}/L_{tot}$--mass
correlations, as well as the iron yields. Without the ICL, the steep slope of 
the $L_{exp}/L_{tot}$--mass correlation ($0.38\pm0.07$, hollow points) shows an 
increasing need for an extra light component (or more efficient iron production)
in more massive clusters.

ICL may also help with the ``iron excess'' found in clusters with
cool cores. The radial distribution of the iron abundance is 
distinctly separated into two forms: (a) a nearly flat distribution for the
clusters without a cool core (NCC-- non-cool core clusters), and (b) a centrally
peaked profile for the cool core (CC) clusters. The amount of iron from the 
central part of CC clusters above the uniformly distributed iron
floor (seen also in NCC clusters) is called the iron excess.  Using a sample of 
12 CC clusters, DG04 examine the relation between the BCGs and the iron
excess in the clusters. Using standard stellar population synthesis models, they
find the extra amount of iron is comparable to the iron in supernova ejecta from
the BCGs.

Here we evaluate the role of ICL (and BCG) in this process in a very simple way;
given the excess iron mass in a cluster, we calculate the expected amount of 
light in the core, if the IMLR for the excess iron is to be consistent with a SSP model.  We have the BCG luminosity for 10 of 12 CC clusters in the DG04 sample (excluding A2142), 
and we again use Eqn~\ref{eq:icl} to estimate the amount of ICL
within the central $0.2 r_{200}$ of the clusters (assuming the ICL is 
distributed similarly to the galaxies, namely an \citealt{navarro97} profile 
with concentration of 3, see \S 3.1 in paper II). We use 
the value of the iron excess provided in Table 1 of DG04. 
The ratio between the SSP IMLR-inferred total light (Salpeter IMF) and our 
ICL+BCG light is shown as stars in the lower panel of Fig~\ref{fig:iron}. It is
clear that for some clusters, the BCG alone may account for the iron excess,
but in other cases ICL helps to provide sufficient iron. Repeating the calculations for
the Kroupa IMF, we find that on average $\overline{L_{exp}/(L_b+L_i)}=2.1
\pm0.3$.  
Assuming a more concentrated radial profile for the ICL (e.g. the de Vaucouleurs
profile found by \citealt{gonzalez04}), we find that $0.2 \le
\overline{L_{exp}/(L_b+L_i)} \le 0.9$ for the Salpeter IMF.
It it clear that the ICL reservoir of stars is helpful in explaining the iron excess in cool core clusters.

To summarize, we find that: (1) the ICL lowers the mean iron yield per solar mass of stars for the 
cluster iron production, although the value is still high compared to that of
the solar vicinity; (2) it is possible to generate the amount of iron observed in clusters with the yields
consistent with simple stellar population models, if a Salpeter IMF is assumed and an ICL amount 
predicted by our model is present;  (3) the ICL makes the 
production efficiency of iron more uniform for clusters of different masses,
which is reflected in the near-flat iron yields and the flat $L_{exp}/L_{tot}$ 
ratios, with respect to cluster mass; (4) the ICL can help in the production of the excess iron in the cores of CC clusters (by lowering the 
yields and shortening the enrichment time, c.f. \citealt{boehringer04}).  The ICL not only helps produce the metals, but also circumvents the problem of transporting the metals out of galaxy halos and into the ICM.  However, if the IMF is consistent with that observed in the solar vicinity, simple stellar population models require more than twice as much more light to
explain the iron production in clusters. 

We note that the above calculations only deal with the {\it amount}
of iron that can be produced, but not the relative abundances between iron and
other elements. More sophisticated models that also invoke an ICL component are
needed to fully investigate the ICM enrichment process.

\section{Discussion}
\label{sec:disc}

\subsection{Choice of Photometry}

As a test of the robustness of our results, we conduct the analysis presented 
in \S\ref{sec:bcglite} using the 2MASS
extrapolated total magnitudes for the BCGs. We find that it makes no
difference on the $L_b$--$M$ relation whether isophotal (corrected by 0.2 mag) or total magnitudes 
are used.

In this paper we choose not to examine the BCG light from magnitudes measured at
a fixed metric radius, as many investigators do \citep[e.g.][]{schneider83b,brough02}. The 
reason is that the sizes of BCGs are not fixed; we find that the half-light 
radius of the BCGs
correlates with mass of the host cluster. Using the isophotal magnitudes has the
advantage of reflecting this correlation, which may provide some insights into 
the BCG formation histories.

\subsection{BCG Sample Selection}
\label{sec:usgc}

Our cluster sample is basically \xray selected, which may bias against those
loose, low mass systems whose \xray emission is weak. 
In order to assess the effects of this possible bias on the BCG properties, we 
investigate the properties of the BCGs in a large
sample of optically selected groups and clusters (the UZC-SSRS2 group catalog, 
hereafter UZC groups, \citealt{ramella02}). For each member galaxy we search in 
the 
2MASS extended source catalog its $K$-band counterpart within $25\arcsec$ of the
position listed in the catalog. We focus on groups of five or more members
that are distinct from our sample clusters and groups. 
Note that only $40-80\%$ of the UZC groups with five or more members are 
estimated to be real physical systems \citep{ramella02}.
The virial mass for each group is from Table 1 of the catalog \citep{ramella02}.

In Fig~\ref{fig:gbcg} we compare the properties of these BCGs with those in our 
sample.  The upper panel shows the $L_b$--$M$
correlation (the crosses are 291 UZC groups, the squares are the
UZC groups that have detected \xray emission from the {\it ROSAT} All-Sky 
Survey, \citealt{mahdavi00}, and the shaded area show the distribution of BCGs 
in our sample). We see
that the X--ray-bright UZC groups seem to occupy the same regions in the
$L_b$--$M$ space, if only somewhat less massive, while the X--ray-faint groups
exhibit large scatter in both BCG luminosity and system mass.
Possible sources for the scatter include: (1) $20-60\%$ of these systems may
be chance projections (i.e. not real); (2) uncertainties in the estimated 
virial mass; 65\% of the systems have only five members; we note that with only 
5 (10) redshift measurements, the fractional 1$\sigma$ uncertainty in mass is about 95\% 
(67\%). Even in the
limit that a large number of redshifts is available, the dynamically inferred
mass may still be affected by the presence of substructure or surrounding large
scale structure. A comparison between X--ray-derived mass ($M_X$) and velocity 
dispersion-inferred mass ($M_V$) for 8 massive ($M_{X,200}\ge5\times10^{14}
M_\odot$) clusters in common between our sample and the UZC catalog shows that 
the average mass difference is $\overline{M_V/M_X}=6.5$ (number of redshifts
based on which dynamical masses are derived ranges from 19 to 144; the smallest
and largest $M_V/M_X$ ratios are 1.9 \& 12.3, which are from 144 \& 72 
redshifts, respectively). Although there is a
correlation between the number of redshifts measured and the mass difference,
some outliers that strongly deviate from the correlation exist.
In our analysis we regard the \xray emission weighted temperature as a more 
reliable mass estimator.

Under the assumption that the clusters are largely regular systems, at a given
mass we expect a cluster to contain a certain number of galaxies (e.g. from the 
observed $N$--$M$ relation).
We therefore can select only groups whose number of members is at least 10\% of 
the expected galaxy number based on our observed $N(M)$ relation; with this
requirement, for a given BCG luminosity, clusters in our sample represent the
most massive systems (i.e. crosses that lie on the lower right part of the 
shaded area will be removed by this criterion).

Comparing the BCGs in the X--ray-bright UZC systems (squares)
with those in our sample (shaded region), we see that
there is no suggestion of flattening of the $L_b$--$M$ correlation below 
$10^{14}M_\odot$ (c.f. \S\ref{sec:bcglite}).  For a given cluster mass, this 
comparison shows how much scatter there may be for the luminosity of BCGs in the
local universe. 

The lower panel of Fig~\ref{fig:gbcg} shows the BCG light fraction for a subset
of the UZC groups (crosses). We only show those systems which lie close enough so
that the 2MASS probes down to or below $M_K=-21$. These systems seem to form a 
continuation from our sample (solid points) in their BCG light fraction. This is
interesting, because the two samples are selected very differently (optical vs X-ray); however, we caution again that mass uncertainties in the low mass UZC groups may be 
large. The crosses surrounded by a square show the groups with ten
or more redshift measurements. The BCG light fraction in these groups is indeed
consistent with the expectations from our sample.  We infer that the galaxy 
formation process is highly regular, and that the halo mass is the single most 
critical parameter in determining the BCG light fraction.

\begin{inlinefigure}
   \ifthenelse{\equal{\figtype}{EPS}}{
   \begin{center}
   \epsfxsize=8.cm
   \begin{minipage}{\epsfxsize}\epsffile{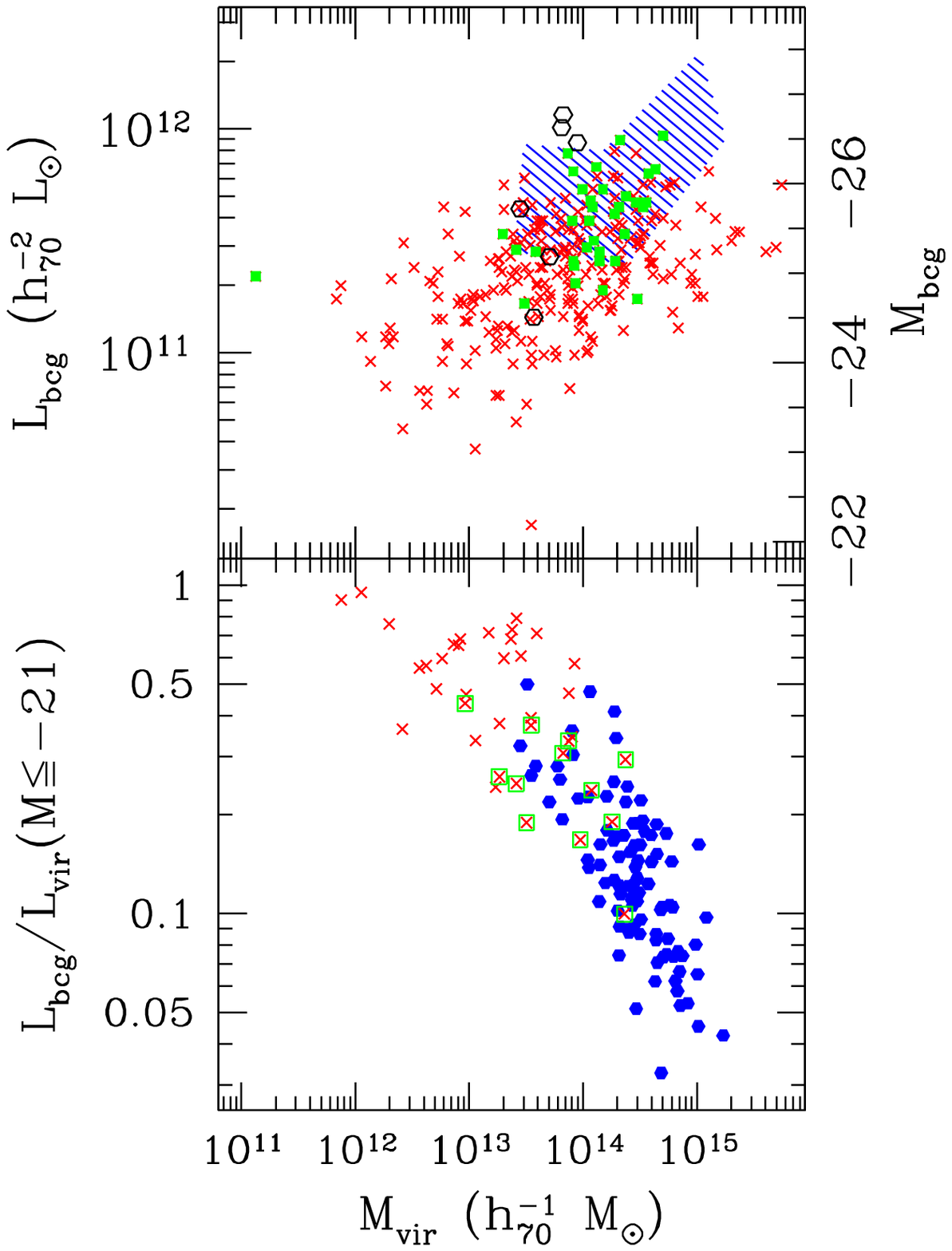}\end{minipage}
   \end{center}}
   {\myputfigure{f10.pdf}{1.9}{1.7}{-130}{-90}}
   \figcaption{\label{fig:gbcg}
        Upper panel: the BCG luminosity for the UZC systems (the crosses) and
	for our sample (shaded area); those UZC systems with detected \xray
	emission are denoted by squares. (Circles: the ``fossil group'' BCGs,
	see \S\ref{sec:fossil}.)
	Lower panel: the BCG contributions to 
	the total light emitted from galaxies. The crosses are the UZC BCGs,
	the crosses with a square are the UZC groups with ten or more redshift
	measurements,
	the solid points are BCGs in our sample. We only show the UZC systems
	that are probed down to at least $M_K=-21$ by the 2MASS.
     }
\end{inlinefigure}

\subsection{Fossil Groups}
\label{sec:fossil}

There exists a class of galactic systems with masses comparable to groups of
galaxies, with diffuse \xray emission from hot gas, but where
the optical light is totally dominated by one single galaxy (the so-called 
``fossil groups'', e.g. \citealt{jones03} and references therein). One property
for a group to qualify as a fossil group is that the (optical) magnitude difference 
between the brightest and second ranked galaxies exceeds $\Delta m_{12}\ge 2$.
This in turn implies that the BCG luminosity fraction in these groups is large
(e.g. 70\%, \citealt{jones03}, hereafter J03). Based on an X-ray flux-limited sample, 
J03 suggest that the fossil groups constitute $8-20\%$ of all systems with 
comparable \xray luminosity ($L_X\ge 5\times 10^{41} h_{70}^{-2}$ erg/sec), and
may be an important source for producing the BCGs in more massive systems (after
the groups fall into larger systems).  We show in Fig~\ref{fig:bcglf} that in 
the lowest mass systems in our sample, the 
BCG light fraction approaches 50\%. Our merger picture of BCG growth also 
implies that the BCGs in lower mass systems serve as building blocks for the
brightest galaxies in more massive systems. 

To examine whether these fossil groups are different from our low mass
groups, we identify the six fossil group dominant galaxies listed in 
\citet{jones00,
jones03} in the 2MASS extended source catalog, and estimate the group mass from
either \xray luminosity or temperature. The masses of these systems are 
estimated to lie in the range $2-9 \times 10^{13} M_\odot$ (again, we caution
large uncertainties in mass). Compared to our BCGs
and those in the UZC groups, half of the fossil group galaxies show extraordinary
luminosity in the sense that they are $0.3-0.5$ mag brighter than expected 
for the mass of their host groups (Fig~\ref{fig:gbcg}, upper panel, hollow circles). 
The other half appear at normal or even dimmer luminosity within the mass range 
(we note, however, that no lower limit on the BCG optical luminosity is imposed
in J03's criteria for fossil groups).
Due to the depth of 2MASS, it is not possible to
further examine the luminosity fraction of these galaxies in $K$-band. 
However, there is one system in each of our sample and the UZC catalog that 
appears to be a fossil group: AWM04 \& U085. Both have $\Delta m_{12}>2$ in 
$K$ and are X-ray luminous. The BCG luminosities are $7.1\times10^{11}h_{70}^{-2}L_\odot$
\& $7.8\times10^{11}h_{70}^{-2} L_\odot$, and the BCG light fractions are
0.41 \& $<0.63$,\footnote{Note that the total light in U085 is estimated to 
$M_K=-22.1$ only. We have ignored the criterium that $\Delta m_{12}\ge2$ is
to be applied for galaxies within half of the virial radius (J03).} 
respectively. The BCGs of these two systems are slightly more luminous than the
expectation from Eqn~\ref{eq:bcglm}, and the BCG luminosity fractions do not
show a total dominance. Based on only few systems, it is not possible to
draw definitive conclusions. However, it is suggestive that the fossil groups
found in J03 are a natural extension in the very low mass regime of the BCG behavior
found for the large cluster sample studied in this paper.
Larger cluster samples and deeper photometry will be needed to better examine
this possibility.

\subsection{BCG Formation Scenarios}
\label{sec:bcgformdisc}

In \S\ref{sec:bcgform} we discuss some possible BCG evolution scenarios.
Currently both observational and numerical studies favor the theory that BCGs
form rapidly from mergers of several galaxies during an early epoch of cluster or
group collapse, followed by a rather limited subsequent growth in luminosity
\citep[e.g.][]{merritt84,lauer88,tremaine90,dubinski98,oegerle01}. 
We argue, however, that BCGs have to merge with other 
galaxies, most likely the BCGs from merging groups, as they are the most
efficient source of light for the BCG light growth. Support for this picture
includes the fact that (1) moderately luminous ($\sim L_*$) galaxies that are
more abundant in low mass clusters do not contain enough light
to account for the growth in BCG luminosity, (2) there are enough very luminous 
galaxies available to account for the BCG luminosity growth, (3) massive galaxies 
can indeed merge within relatively short timescales \citep{tremaine90}
and (4) massive galaxies are more likely to merge because of the larger cross
sections (or gravitational focusing) and shorter dynamical friction timescales \citep[e.g.][]{klypin99}.

In such a scenario, the BCG co-evolves with the cluster. The tight correlation 
between the BCG light fraction
and cluster mass strongly supports this view. The scatter
in the BCG light on group scales may reflect the heterogeneous origin of these
galaxies (c.f. Figs~\ref{fig:bcglm}, \ref{fig:gbcg}); if they were the outcome of mergers among several massive galaxies
during cluster collapse, we do not expect their properties to be as
homogeneous as those in high mass clusters. However, as these building blocks
merge and form more massive BCGs, the mixing makes their properties more
homogeneous. 

One possible consequence of the merger is 
that BCGs in more massive clusters may have different structure than
those in low mass systems. We find that the half-light radius $r_e$ of the BCGs
positively correlates with cluster mass. Deep photometry, which is needed for 
examination of other structural parameters, such as $\alpha \equiv d\log L/ d\log r|_{r_m}$, 
the logarithmic luminosity derivative at a fixed metric radius $r_m$ (e.g. \citealt{hoessel80b,postman95},
however, see \citealt{collins03}), or spectroscopic studies that reveal the
stellar kinamatics of the BCGs to large radii (e.g. \citealt{weil96}), may 
further test this evolution scenario. 
Although we expect these mergers to be mainly
without star formation, color information may provide valuable clues.

Our suggestion for BCG light (mass) growth comes from the clusters in the nearby
universe; by studying the BCGs in clusters of different cosmic epoch, 
constraints have
also been placed on BCG growth \citep[e.g.][]{aragon98,collins98,burke00,
nelson02}. It has been suggested that BCGs in X--ray-luminous clusters seem to
evolve passively, with no or little (stellar) mass accretion, while those in
less luminous clusters seem to exhibit strong mass accretion, or no evolution
in luminosity \citep[e.g.][and references therein]{brough02,nelson02}.
We note that using a fixed \xray luminosity as a cluster mass division
is not ideal, because at larger redshifts the same luminosity corresponds to
progressively smaller mass.  Comparison of BCGs in low
\xray luminosity clusters across a wide range in redshift introduces a mass trend which
can explain the inferred strong BCG mass accretion with decreasing redshift.
As for \xray luminous clusters, the resolution may lie in the way the $L_b$--$M$
correlation evolves with time.

This also leads to the issue noted in \S\ref{sec:bcgform}: if the BCGs of 
different masses do form
a continuum in their evolution, the progenitors of BCGs in present-day high
mass clusters are not necessarily the {\it present-day} low mass cluster BCGs. It is
BCGs in low mass clusters at higher redshifts that should be compared to the
present-day high mass cluster BCGs.  We will investigate this in a future 
publication.

\section{Summary}
\label{sec:summary}

Based on the rich dataset provided by the 2MASS, we systematically examine 
various cluster 
properties in the near-IR $K$-band for a large sample of 93 clusters and groups
(papers I \& II). Paper I develops the basic technique. We study the
correlation between total galaxy light and cluster binding mass, and the halo
occupation distribution in paper II. Here we focus on the extraordinary 
properties of the brightest cluster galaxies, and their implications for the
total cluster light budget and for cluster evolution.
The main findings of this paper are summarized as follows:

1. The BCG projected position coincides with the peak in \xray emission,
in agreement with previous findings. In $\sim 80\%$ of the cases, the BCG
location can serve as cluster center to within 10\% of the virial radius.

2. The $K$-band BCG luminosity shows clear correlation with cluster mass; 
$L_{b} \propto M^{0.26\pm 0.04}$ for all clusters in our sample.
Combined with a large optically-selected group sample (from the UZC-SSRS2 
catalog), we find that the correlation between the BCG luminosity and the 
cluster mass extends to even lower mass scales.

3. We argue that BCGs likely grow in luminosity via mergers with other
luminous galaxies, most likely BCGs from subclusters that have fallen into the 
cluster. A 
comparison between the luminosity functions in high and low mass clusters shows a lower density of luminous galaxies ($\sim L_*$) in more massive clusters.  The luminosity from these missing galaxies could not make up the differences between the BCG luminosity in high and low mass clusters.  On the other hand, based on the luminosity distribution of clusters at different
masses, there appear to be enough very luminous galaxies ($>L_*$) in
intermediate or low mass clusters to supply the luminosity needed for the BCG 
growth with increasing cluster mass.

4. The BCG-to-total galaxy light ratio is a decreasing function of cluster mass;
at group scales BCG light constitutes roughly half of the total light in 
galaxies,
while in the most massive systems in our sample they only account for $5-10\%$ 
of the light in galaxies. The decreasing importance of BCGs in the overall cluster
light budget can be explained if the luminosity growth rates of BCGs (by merging
with other luminous galaxies) are slower than the luminosity growth rates of clusters (by
accreting isolated galaxies or from smaller clusters that have merged).

5. Second- and third-ranked galaxies show clear trends of increasing galaxy
luminosity, but decreasing galaxy-to-cluster light fraction, as a function of 
cluster mass, as do the BCGs. These
similarities suggest that the brightest galaxies share a similar
formation and evolution history.  At group scales the few brightest galaxies 
make up almost all the light, while at the other mass extreme
the brightest three galaxies only account for $10\%$ of the massive cluster light.

6. We estimate the amount of diffuse light that is present in the intracluster 
space (which includes any envelope surrounding the BCG) as a function of cluster mass.
Two simple models are considered; the first one makes use of the observed total
galaxy light--cluster mass relation, as well as the BCG light--cluster mass 
correlation. The second model takes into account the assembly history of 
clusters, and uses the observed $L$--$M$ relation. Both models suggest that the 
amount of ICL increases with cluster mass; for clusters more massive than
$10^{14} M_\odot$ the ICL light fraction ranges from 30\% to 60\%. The
model predictions are in reasonable agreement with both direct observations and
numerical simulations. 

7. The cluster baryon fraction, including the ICL (roughly a 100\% increase in cluster stellar mass), 
is in good agreement with the  {\it WMAP} result \citep{bennett03}.  The ICL component reduces the discrepancy between the observed value of the cold baryon fraction ($\sim16$\%) and that found in numerical simulations.

8. The ICL reduces the cluster iron yields by $\sim 40\%$, to a value of 3 
times solar, still high compared to the solar vicinity. However, with the ICL included, the observations indicate a uniform iron production efficiency for clusters of different masses. Within a
simple stellar population model described by a Salpeter IMF, the observed starlight in galaxies and that implied for the ICL is high enough to explain the high iron abundance in clusters.   
The ICL also helps in accounting for the production of the observed ``iron excess'' 
found in the cool-core clusters.

\acknowledgements

We acknowledge A. Sanderson for help on obtaining relevant information of 
BCGs from the NED.  
We thank A. Finoguenov, A. Gonzalez, P. Ricker, C. Sarazin, 
F. van den Bosch, and D. Zaritsky for useful discussions and suggestions.
We thank an anonymous referee for helpful comments.
This work was supported in part by the NASA Long Term Space Astrophysics 
award NAG 5-11415.  This publication makes use of data products from the Two 
Micron All
Sky Survey, which is a joint project of the University of Massachusetts and the
IPAC/Caltech, funded by the NASA and the NSF. This research has made use of the
NED and BAX.  

\bibliographystyle{apj}
\bibliography{cosmology,refs}

\end{document}